\documentclass[12pt]{article}
\usepackage{amsmath}
\usepackage{graphicx,psfrag,epsf}
\usepackage{float}

\usepackage{color}
\usepackage{scalefnt}
\usepackage{multirow}
\usepackage{enumerate}
\usepackage{natbib}
\usepackage{url} % not crucial - just used below for the URL 
\usepackage{amsthm,amsfonts,amssymb,amsxtra}%,empheq}
\usepackage[T1]{fontenc}
\usepackage{array,booktabs}%,arydshln}

\newcommand{\bfbeta}{\mbox{\boldmath{$\beta$}}}
\newcommand{\bfvarepsilon}{\mbox{\boldmath{$\varepsilon$}}}

\newcommand{\bfmu}{\mbox{\boldmath{$\mu$}}}

\newcommand{\calX}{\mathcal{X}}

\usepackage{xr}

\newcommand{\blind}{0}

\begin{document}

\def\spacingset#1{\renewcommand{\baselinestretch}%
{#1}\small\normalsize} \spacingset{1}

%%%%%%%%%%%%%%%%%%%%%%%%%%%%%%%%%%%%%%%%%%%%%%%%%%%%%%%%%%%%%%%%%%%%%%%%%%%%%%

\if0\blind
{
  \title{\bf Prediction properties of optimum response surface designs}
   \author{Heloisa M. de Oliveira\thanks{
    The authors gratefully acknowledge financial support from Coordena\c{c}\~ao de Aperfei\c{c}oamento de Pessoal de N\'ivel Superior (PNPD/CAPES, Brazil Government) and from FAPESP grant numbers 2013/09282-9 and 2014/01818-0.}\hspace{.2cm}\\
    Universidade Federal de Santa Catarina, Curitibanos, SC, 89520-000, Brazil\\
    C\'esar B. A. de Oliveira\\
    Instituto Tecnol\'ogico de Aeron\'autica, S\~ao Jos\'e dos Campos, SP, 12228-900, Brazil\\
    Steven G. Gilmour\\
    King's College London, London, WC2R 2LS, UK\\
    and \\
    Luzia A. Trinca\\
   Universidade Estadual Paulista, Botucatu, SP, 18618-689, Brazil}
  \maketitle
} \fi

\if1\blind
{
  \bigskip
  \bigskip
  \bigskip
  \begin{center}
    {\LARGE\bf On prediction properties of optimum response surface designs}
\end{center}
  \medskip
} \fi

\bigskip

\begin{abstract}
Prediction capability is considered an important issue in response surface methodology. Following the line of argument that a design should have several desirable properties we have extended an existing compound design criterion to include prediction properties. Prediction of responses and of differences in response are considered. Point and interval predictions are allowed for. Extensions of existing graphical tools for inspecting prediction performances of the designs in the whole region of experimentation are also introduced. The methods are illustrated with two examples.
\end{abstract}

\noindent%
{\it Keywords:} compound criteria, dispersion graphs, $(DP)$-optimality, FDS, $I$-optimality, pure error.
\vfill

\newpage
\spacingset{1.45} % DON'T change the spacing!
\section{Introduction}
\label{sec:intro}

Experiments provide important information for discoveries in many research areas. 
Careful planning of an experiment is very important in order to obtain informative answers to the questions of the research problem at hand. The planning 
phase can be quite involved and methods for finding optimum designs are very useful when there are several quantitative factors related to the response variables of interest and when there are practical restrictions. Work in this area started by considering the optimization of single design-criterion functions aimed at maximizing the precision of the model parameter estimates or prediction of responses. Computational algorithms are well developed mainly 
for $D$- and $I$-efficiency \citep{CookNachtsheim89, jonesgoos2012}. Designs obtained by such methods 
are the best or very close to the best (as they are based on heuristics), given the assumed model, for the property being optimized.  However, for practical purposes, an experiment should answer several research 
questions and so requires a good design with respect to many properties as advocated by \cite{Box-Draper:1975}. Fortunately, in the last decade or so, design 
methodologies seem to be moving in this direction through the application of compound criteria and multiple objective approaches \citep{goosetal2005, jonesnash2011, luetal11, smucker2012, gilmourtrinca2012, smucker2015, borrotti2016, daSilvaGilmourTrinca2017, trincagilmour2017}.

While the use of compound criteria  or multiple objective procedures allow the consideration of a set of one-dimensional properties for constructing the design, graphical techniques add information to illustrate
the prediction properties of the designs. The study of design prediction capabilities through graphs advanced with \cite{g-j&m1989} and \cite{myers1992} when they introduced variance dispersion graphs. These graphs were followed by the quantile plots of 
\cite{khuri96}, the difference variance dispersion graphs of \cite{trincagilmour1999} and the fraction of design space plots of \cite{zahran2003} and \cite{jang2012}. Such techniques are of great value for choosing a final design among many options. 

In this paper we consider a flexible compound criterion for optimization of parameter estimation properties as well as
prediction. The paper introduces several new methods, namely: (i) difference fraction of design space plots, which show variances of differences in response; (ii) variance dispersion graphs and fraction of design space plots for interval predictions, for both responses and differences in response; (iii) the $I_D$ criterion, for point estimation of differences in response; (iv) the $(IP)$ and $(I_DP)$ criteria for interval estimation of responses and differences in response; (v) using standard errors, rather than variances in the plots; (vi) using relative volume in the plots. These methods can be considered as extensions for prediction criteria motivated by the difference variance dispersion graphs of \cite{trincagilmour1999} and the adjusted criteria of \cite{gilmourtrinca2012}. The designs constructed are further evaluated according to their performances with respect to prediction capabilities using the graphs described and extensions incorporating the new measures. In Section \ref{sec:DC} we
review the literature and propose extensions to the usual design criteria. In Section \ref{sec:PC} we discuss graphical methods for prediction evaluation and propose two
extensions, and in Section \ref{sec:Ex} we illustrate these methods and compare several designs for two examples. Motivated by these results, we note in Section \ref{sec:CCD} some situations in which central composite designs are optimal. Finally a discussion is presented in Section \ref{sec:disc}.

\section{Design criteria}\label{sec:DC}
Data from experiments with $q$ continuous quantitative factors are routinely analyzed by fitting low order polynomials. These are used as approximations to the
unknown true function relating the response variable $Y$ and the treatments. A treatment $\mathbf{x}$ is defined by a specific combination of levels of the $q$ factors
$X_1,~X_2,~\ldots,~X_q$. The full model for a completely randomized design with $n$ experimental units (runs) is 
\begin{equation}
\mathbf{Y} = \bfmu(\mathbf{x})+\bfvarepsilon, 
\label{eq:full}
\end{equation}
where $\mathbf{Y}$ is the column vector of random variables of dimension $n$, $\bfmu(\mathbf{x})$ is the mean vector of $\mathbf{Y}$, depending on $\mathbf{x}$, and $\bfvarepsilon$ is the error term random vector satisfying $E(\bfvarepsilon)=\mathbf{0}$ and $Var(\bfvarepsilon)=\sigma^2\mathbf{I}$. 
The full model may be further approximated by 
\begin{equation}
\bfmu(\mathbf{x})\approx\mathbf{X}\bfbeta,
\label{eq:poly}
\end{equation} 
where, using standard notation, 
$\bfbeta$ is the $p$-dimensional vector of unknown parameters and $\mathbf{X}$ is the $\big(n \times p \big)$ model matrix whose rows, denoted by $\mathbf{f}(\mathbf{x})^\prime$, 
are expansions of levels of the factors in order to accommodate the desired polynomial. 

Since the matrix $\mathbf{X}$ is defined by the design and the model approximation, for notational simplicity we will refer to the design as $\mathbf{X}$.
As discussed in \cite{gilmourtrinca2012}, fitting the full model (\ref{eq:full}) allows unbiased estimation of $\sigma^2$ if degrees of freedom from treatment 
replications are available while fitting model (\ref{eq:poly}) allows simplification and also lack of fit checking if there are spare treatment degrees 
of freedom. In order to construct optimum designs that allow unbiased estimation of error variance, \cite{gilmourtrinca2012} proposed adjustments to the usual alphabetical design criteria,
based on the appropriate quantiles of the $F$ distribution, e.g.\ the $(DP)_S(\alpha)$ and $(AP)_S(\alpha)$ criteria. Following their logic, Goos, in the discussion of \cite{gilmourtrinca2012} proposed the same type of adjustment 
for the $I$-optimality criterion.

\subsection{Prediction of responses}

For any point $\mathbf{x} \in \calX$, $\calX$ being the region which the experimenter desires to explore, the variance of $\hat{y}(\mathbf{x})$, the estimated response from the fitted polynomial, 
is $\text{var}(\hat{y}(\mathbf{x}))=\sigma^2 \mathbf{f}(\mathbf{x})^\prime(\mathbf{X}^\prime\mathbf{X})^{-1}\mathbf{f}(\mathbf{x})$. An $I$-optimum design $\mathbf{X}$ is such 
that the average variance of predictions over the whole experimental region $\calX$ is minimized. Let $\Psi=\int_{\mathbf{x}\in\calX}d\mathbf{x}$ be the volume of the region 
$\calX$. The average prediction variance is defined as
\begin{equation}
\text{average}~\text{variance}=\Psi^{-1}\int_{\mathbf{x}\in\calX}\text{var}(\hat{y}(\mathbf{x}))d\mathbf{x}\propto\int_{\mathbf{x}\in\calX}\mathbf{f}(\mathbf{x})^\prime(\mathbf{X}^\prime\mathbf{X})^{-1}\mathbf{f}(\mathbf{x})d\mathbf{x}.\label{eq:vary}\end{equation} As the integrand in (\ref{eq:vary}) is a scalar, and using properties of the trace of matrix products, it is easily shown that 
\begin{equation}
\text{average}~\text{variance}\propto\text{trace}\left[\mathbf{\mathcal{M}}(\mathbf{X}^\prime\mathbf{X})^{-1}\right],\label{eq:I}\end{equation}where $\mathbf{\mathcal{M}}=\int_{\mathbf{x}\in\calX}\mathbf{f}(\mathbf{x})\mathbf{f}(\mathbf{x})^\prime d\mathbf{x}$
is the so called moment matrix of the region. For regular spherical and  cubic regions and polynomial models, the matrix $\mathcal{M}$ obeys known patterns, given explicitly, for the full second order model, 
in \cite{hardin1991sphere} and \cite{hardin1991cube} for example. 

Considering that interest is in evaluating the performance of the design for interval predictions, the $I$ criterion may be modified to minimize the average, over the design region $\calX$, of the width of pointwise confidence intervals for the mean response. This gives the criterion function
\begin{equation}
\text{trace}\left[\mathbf{\mathcal{M}}(\mathbf{X}^\prime\mathbf{X})^{-1}\right]F_{1,d;1-\alpha_3},\label{eq:IP}\end{equation} the $(IP)(\alpha_3)$ criterion, where $d$ is the number of 
pure error degrees of freedom of the design $\mathbf{X}$, $1-\alpha_3$ is the confidence level for pointwise intervals for $E(y(\mathbf{x}))$ and $F_{1,d;1-\alpha_3}$ is the relevant quantile from the $F$ distribution. According to several researchers, prediction is a key point for planning  
response surface experiments \citep{g-j&m1989, hardin1993, trincagilmour1999, zahran2003, goosjones2011book, jonesgoos2012, borrotti2016}. 

\subsection{Prediction of differences in response}

In \cite{trincagilmour1999} it was argued that rather than the response level, prediction of differences in 
responses would be more interesting. In particular, we are often interested in differences between the estimated response at the expected optimum or standard operating conditions and the estimated response at other 
locations, i.e.\ $y(\mathbf{x})-y(\mathbf{x}_0)$, where $\mathbf{x}_0$ denotes standard conditions or the prior expected optimum combination. We code the factors, so that $\mathbf{x}_0=\mathbf{0}$, which implies that the focus should be on estimating $y(\mathbf{x})-\beta_0$. There are both theoretical and practical reasons why predicting differences in response makes more sense than predicting responses themselves.

First, the randomization of the experiment ensures that least squares estimators of the parameters are unbiased, except for the estimate of $\beta_0$, which requires the further assumption that the experimental units are a random sample from a population of possible units - see for example \cite{coxreid}, p.32-36, or Chapter 5 of \cite{hinkelmann}. In response surface studies the runs are almost never a random sample and even treating them as a representative sample is usually implausible. Therefore predictions of responses made from the experiment cannot reasonably be applied to the process over time, but predictions of differences in response can.

Secondly, important aspects of the interpretation of fitted response surfaces, such as estimating the location of the stationary point and estimating the location of ridges, do not depend on the intercept. For example, the stationary point is located at $-\mathbf{B}^{-1}\mathbf{b}/2$, where $\mathbf{b}$ and $\mathbf{B}$ contain respectively the first and second order parameters. Similarly, canonical analysis depends on the same vector and matrix. Thus important aspects of response surface interpretation, which are difficult to build directly into design optimality criteria, should be better represented by optimizing the prediction of differences in response than by optimizing predictions of responses.

Finally, if $\mathbf{x}_0$ represents standard operating conditions of the process, we should already have a much better estimate of $E[y(\mathbf{x}_0)]$ from the historical running of the process than we can expect to get from a fairly small experiment. Using the factor coding, we can treat this historical estimate as being the true $\beta_0$. Then the best prediction from the experiment of the response at some $\mathbf{x}$ is not $\hat{y}(\mathbf{x})$, but 
\begin{equation}
\label{eq:ytilde}
\tilde{y}(\mathbf{x}) = \beta_0 +\hat{y}(\mathbf{x}) -\hat{\beta}_0.
\end{equation} 
Then the variance of a prediction using this method is
\[
\text{var}[\tilde{y}(\mathbf{x})] = \text{var}[\hat{y}(\mathbf{x}) -\hat{\beta}_0] = \text{var}[\hat{y}(\mathbf{x})-\hat{y}(\mathbf{x}_0)].
\]
Hence, even if predictions of responses are of interest, the design should be chosen to minimize variances of differences in response.

Based on this argument, we define the $I_D$ criterion which minimizes the average difference variance,
\begin{eqnarray}\text{average}~\text{difference}~\text{variance}&=&\Psi^{-1}\int_{\mathbf{x}\in\calX}\text{var}[\hat{y}(\mathbf{x})-\hat{y}(\mathbf{x}_0)]d\mathbf{x} \nonumber\\
&\propto&\int_{\mathbf{x}\in\calX}[\mathbf{f}(\mathbf{x})-\mathbf{f}(\mathbf{x}_0)]^\prime(\mathbf{X}^\prime\mathbf{X})^{-1}[\mathbf{f}(\mathbf{x})-\mathbf{f}(\mathbf{x}_0)]d\mathbf{x}.\end{eqnarray}
For coded factors $\mathbf{x}_0=\mathbf{0}$ and analogously to (\ref{eq:I}) we have \begin{equation}\text{average}~\text{difference}~\text{variance}\propto\text{trace}\left[\mathbf{\mathcal{M}}_0(\mathbf{X}^\prime\mathbf{X})^{-1}\right],\end{equation}
where $\mathbf{\mathcal{M}}_0=\int_{\mathbf{x}\in\calX}[\mathbf{f}(\mathbf{x})-\mathbf{f}(\mathbf{0})][\mathbf{f}(\mathbf{x})-\mathbf{f}(\mathbf{0})]^\prime d\mathbf{x}$ such that $\mathbf{\mathcal{M}}_0$ is the $\mathbf{\mathcal{M}}$ matrix with first row and first column set to zero. Similarly to the $(IP)(\alpha_3)$ criterion we may now define the $(I_DP)(\alpha_{4})$ criterion that searches for $\mathbf{X}$ which minimizes \begin{equation}
\text{trace}\left[\mathbf{\mathcal{M}}_0(\mathbf{X}^\prime\mathbf{X})^{-1}\right]F_{1,d;1-\alpha_{4}},\label{eq:I_DP}\end{equation}
where $(1-\alpha_{4})$ is the confidence level for pointwise intervals for expected response differences and $F_{1,d;1-\alpha_{4}}$ is the appropriate $F$ distribution quantile. This minimizes the average, over the design region $\calX$, of the width of pointwise confidence intervals for the mean response if we use equation (\ref{eq:ytilde}) for the predictions.

\subsection{Compound criteria}

\cite{hardin1993} and \cite{jonesgoos2012} showed that $I$-optimum designs have smaller losses in efficiency for parameter estimates than $D$-optimum designs have in terms of prediction 
efficiency. Whereas these authors preferred $I$-optimality on this basis, it is more desirable to build both parameter estimation and prediction into the optimality criterion. This, together with the commonly accepted view that a design should have several good properties, suggests investigating a compound criterion for prediction as well as 
estimation.  To that end we  extend the compound criteria 
of \cite{gilmourtrinca2012} in order to take into account predictions of the response as well as expected differences in the response with respect to the experimental region center. Thus we simply divide \cite{gilmourtrinca2012}'s equation (5) by 
\begin{equation}
 {F^{\kappa_6}_{1;d;1-\alpha_{3}}} {F^{\kappa_8}_{1;d;1-\alpha_{4}}}\text{tr}\left\{\mathbf{\mathcal{M}}(\mathbf{X}^\prime\mathbf{X})^{-1}\right\}^{\kappa_5+\kappa_6}\text{tr}\left\{\mathbf{\mathcal{M}}_0(\mathbf{X}^\prime\mathbf{X})^{-1}\right\}^{\kappa_7+\kappa_8},
 \end{equation}
where $\kappa_5,~\kappa_6,~\kappa_7~\text{and}~\kappa_8$ are the priority weights for point response prediction, interval response prediction, point response difference prediction and interval response difference prediction, respectively, leading to the more general compound criteria, after ignoring constant terms, given by
\begin{equation}
\frac{F^{-\kappa_1}_{p-1,d;1-\alpha_1}{F^{-\kappa_2}_{1,d;1-\alpha_2}}|\mathbf{X}_0^\prime\mathbf{Q}\mathbf{X}_0|^\frac{\kappa_0+\kappa_1}{p-1}
(n-d)^{\kappa_4}{F^{-\kappa_6}_{1,d;1-\alpha_3}} {F^{-\kappa_8}_{1,d;1-\alpha_4}}}{
\text{tr}\{\mathbf{W}(\mathbf{X}^\prime\mathbf{X})^{-1}\}^{\kappa_2+\kappa_3}\text{tr}\left\{\mathbf{\mathcal{M}}(\mathbf{X}^\prime\mathbf{X})^{-1}\right\}^{\kappa_5+\kappa_6}\text{tr}\left\{\mathbf{\mathcal{M}}_0(\mathbf{X}^\prime\mathbf{X})^{-1}\right\}^{\kappa_7+\kappa_8}},
\label{eq:CP}
\end{equation}
where $\sum_{i=0}^8\kappa_i=1$ and $\mathbf{X}_0$ is the $n\times(p-1)$ matrix equal to the $\mathbf{X}$ matrix except that the column of 1's corresponding to the intercept is 
removed and $\mathbf{Q}=\mathbf{I}-\mathbf{1}\mathbf{1}^\prime/n$ is of dimension $n\times n$. Note that we have included in the formula the $D_S$ criterion. By allowing $\kappa_0>0$ we can use the $D_S$ property to reflect parameter point estimation if desired. Note that the formula allows $L$ type criteria, the $A$ criterion being a particular case. For second-order polynomials we recommend the use of weights through the $\mathbf{W}$ matrix in order to adjust the scale for the different types of parameter in the polynomial, i.e.\ linear, quadratic and interaction parameters. 

To find a compromise design by maximizing (\ref{eq:CP}) we can use any algorithm proposed in the literature for factorial designs, such as point- or coordinate-exchange type algorithms.

\section{Design prediction capability}\label{sec:PC}
Many of the measures proposed for design construction and evaluation, e.g.\ those of the type presented in Section \ref{sec:DC}, are global measures that try to convey in a single number all the information available in the design (see the discussion in \cite{anderson-cookmontg2009}). Depending on the objectives  of the experiment, inspection of only these global measures may not suffice for design choice. This is particularly true for prediction since a design may show a reasonable performance globally by performing extremely well in one portion of the region but badly in another portion that could perhaps be of more interest. Thus, for inspection of design capabilities with respect to prediction, several valuable graphical approaches have been proposed. \cite{g-j&m1989} proposed the variance dispersion graphs (VDGs) that plot the maximum, mean and minimum variances for predictions of the response calculated over various spheres within the region of interest. For a scaled region so that the maximum point is at distance 1 from the center, the radius $r$ varies from 0 to 1. From \cite{g-j&m1989}, for the sphere $U_r$ $(U_r = \{\mathbf{x}: \sum_{i=1}^qx_i^2=r^2\}, r<1)$, the mean, or integrated, variance of predictions is the spherical variance defined by 
\begin{equation}
V^r\propto\Psi^{-1}_r\int_{\mathbf{x}\in U_r}\mathbf{f}(\mathbf{x})^\prime(\mathbf{X}^\prime\mathbf{X})^{-1}\mathbf{f}(\mathbf{x})d\mathbf{x}=\text{tr}\{\mathcal{M}_r(\mathbf{X}^\prime\mathbf{X})^{-1}\},
\end{equation}
where  $\Psi_r = \int_{\mathbf{x}\in U_r}d\mathbf{x}$ and $\mathcal{M}_r$ is the matrix of moments for the region $U_r$. \cite{vining1993} gave Fortran code to calculate and plot the maximum, minimum and average variances, for given radius, against the distance from the center. VDGs allow visualization of prediction stability over the region and prediction performance of the design in a more informative way than single valued measures. For cuboidal regions, average variances are not calculated and the maximum and minimum variances are searched over restricted hyperspheres when their radii extrapolate the hypercube. The VDG methodology was extended for inspection of variances of response differences by the introduction of difference variance dispersion graphs (DVDGs) by \cite{trincagilmour1999}. For the sphere $U_r$, the mean or integrated variance of differences between predictions at two points, $\mathbf{x}\in \mathcal{X}$ and the design center, is defined by 
\begin{equation}
DV^r\propto\Psi^{-1}_r\int_{\mathbf{x}\in U_r}(\mathbf{f}(\mathbf{x})-\mathbf{f}(\mathbf{0}))^\prime(\mathbf{X}^\prime\mathbf{X})^{-1}(\mathbf{f}(\mathbf{x})-\mathbf{f}(\mathbf{0}))d\mathbf{x}=\text{tr}\{\mathcal{M}_{0r}(\mathbf{X}^\prime\mathbf{X})^{-1}\},
\end{equation}
where $\mathcal{M}_{0r}$ is the matrix $\mathcal{M}_r$ with first row and first column set to zero.

Because for each design the VDG and DVDG present three (spherical region) or two (cuboidal region) lines it is difficult to compare more than a very few designs in the same plot. Another drawback of these graphs is that they ignore the relative volume associated with the sphere $U_r$ and may lead to misleading interpretations. The situation is more serious for $q\ge4$. 
A more recently preferred display is the fraction of design space (FDS) plot proposed by \cite{zahran2003}.  The FDS plot shows the variance against the relative volume of the region that has prediction variance at or below a given value. 

The FDS plot can be easily extended to difference fraction of design space (DFDS) plots, that is the fraction of design space for variances of the estimated differences between $\hat{y}(\mathbf{x})$ and $\hat{y}(\mathbf{x}_0)$. The usual method to obtain the information for theses graphs is the one outlined in \cite{goosjones2011book} and we use it to obtain FDS and DFDS plots. A very large sample, of size $N$ points, is taken randomly from $\mathcal{X}$ and $v_j=\mathbf{f}(\mathbf{x}_j)^\prime (\mathbf{X}^\prime\mathbf{X})^{-1}\mathbf{f}(\mathbf{x}_j)$ for FDS or $vd_j=(\mathbf{f}(\mathbf{x}_j)-\mathbf{f}(\mathbf{x}_0))^\prime (\mathbf{X}^\prime\mathbf{X})^{-1}(\mathbf{f}(\mathbf{x}_j)-\mathbf{f}(\mathbf{x}_0))$ for DFDS are calculated for $j=1,~2,~\ldots,~N$ ($\mathbf{x}_0$ is fixed at the desired treatment; here we use, as before, $\mathbf{x}_0=\mathbf{0}$). Then these values are sorted such that $v_{(j)}$ (or $vd_{(j)}$) is in the $j^{th}$ position. The graph is simply the plot of $v_{(j)}$ (or $vd_{(j)}$) against $j/(N+1)$. 

We suggest and use an alternative for VDG and DVDG by replacing the radius or distance from the design center by the relative volume of the region inside the hypersphere formed by each distance, to the whole design region. This is particularly useful because we add information that the FDS does not show, that is in which parts of the region the design has which properties. 

The calculation of the values for constructing VDG, DVDG, FDS and DFDS plots is available in the R package \verb"dispersion" \citep{oliveira_cesar2014}. Versions of theses graphs to explore interval prediction properties are easily obtained by multiplying $v_{(j)}$ or $vd_{(j)}$ by $F_{1,d;1-\alpha}$ for some suitable choice of $\alpha$.

\section{Examples}\label{sec:Ex}
In this section we explore the potential of the proposed compound criteria for constructing designs for two experiments. We focus on $D_S$, $(DP)_S$ and prediction efficiencies for constructing the designs. For interval estimation criteria we used $\alpha=0.05$ throughout. The search procedure uses a point exchange algorithm. We further evaluate the prediction capabilities of the designs using several versions of the graphs described in Section \ref{sec:PC}. In the displays we use the standard error (s.e.) instead of the variance scale in order to discriminate better between designs, since most variances are less than 1. The new proposed plots are presented in the paper while slight variations of the old ones are included the in Supplementary Material.

\subsection{Example 1: Cassava bread recipe}
\cite{esc} performed experiments in order to gain knowledge for a gluten-free bread recipe using cassava flour for people with coeliac disease. One of the experiments used $n=26$ experimental units to study the effects of $q=3$ factors, the amount of powder albumen ($X_1$); the amount of yeast ($X_2$) and the amount of cassava flour ($X_3$). Other ingredients and factors associated with the mixing and baking process were kept constant. The experimental region was the cube defined by $10\le x_1\le30g$, $5\le x_2\le 15g$ and $45\le x_3\le65g$, and the experimenter decided to use a modified central composite design (CCD) with four center runs and the factorial part duplicated. One objective was to estimate optimum quantities of the ingredients based on some organoleptic characteristics and the primary model considered was the second-order polynomial with $p=10$ regression parameters. Note that the full three-level factorial would use 27 runs and would allow no pure error degrees of freedom. Alternative designs for this experiment were given by \cite{gilmourtrinca2012}, using the inference based and compound criteria, and in \cite{borrotti2016}, using the multi-objective algorithm, MS-TPLS, for both sets of properties, $D$, $A$ and $I$ and $D_S$, $A_S$ and $I_D$. 
\begin{table}[H]
	%	\scalefont{0.7}
	\caption{\label{tab:desEx1_1} Alternative designs for Example 1 ($n=26$, $q=3$, $p=10$ in cubic region)}
	%	\vspace{0.2cm}
	\centering
	\renewcommand{\arraystretch}{.7}
	\begin{tabular}{rrrcrrrcrrrcrrrcrrr}%	
		\toprule
		\multicolumn{19}{c}{Design}\\ \multicolumn{3}{c}{4}&&\multicolumn{3}{c}{5}&&\multicolumn{3}{c}{6}&&\multicolumn{3}{c}{7} &&\multicolumn{3}{c}{8} \\		
		\multicolumn{3}{c}{{$I$}  }&&\multicolumn{3}{c}{{$(IP)$}}&&\multicolumn{3}{c}{$I_D$}&&\multicolumn{3}{c}{$(I_DP)$}&&
		\multicolumn{3}{c}{$\kappa_1=\kappa_7=.5$} \\
		
		\cmidrule(lr){1-3}\cmidrule(lr){5-7}\cmidrule(lr){9-11}\cmidrule(lr){13-15}\cmidrule(lr){17-19}
		%		\onehalfspacing 		
		$X_1$&$X_2$&$X_3$ &&$X_1$&$X_2$&$X_3$&&$X_1$&$X_2$&$X_3$ &&$X_1$&$X_2$&$X_3$ && $X_1$&$X_2$&$X_3$ \\ 	\cmidrule(lr){1-3}\cmidrule(lr){5-7}\cmidrule(lr){9-11}\cmidrule(lr){13-15}\cmidrule(lr){17-19}
		-1	&	-1	&	-1	&&	-1	&	-1	&	-1&&-1	&	-1	&	-1	&&	-1	&	-1	&	-1	&&	-1	&	-1	&	-1	\\
		-1	&	-1	&	1	&&	-1	&	-1	&	1 &&-1	&	-1	&	-1	&&	-1	&	-1	&	-1	&&	-1	&	-1	&	-1	\\
		-1	&	1	&	-1	&&	-1	&	1	&	-1&&-1	&	-1	&	1	&&	-1	&	-1	&	1	&&	-1	&	-1	&	1	\\
		-1	&	1	&	1	&&	-1	&	1	&	1 &&-1	&	1	&	-1	&&	-1	&	1	&	-1	&&	-1	&	-1	&	1	\\
		1	&	-1	&	-1	&&	1	&	-1	&	-1&&-1	&	1	&	1	&&	-1	&	1	&	1	&&	-1	&	1	&	-1	\\
		1	&	-1	&	1	&&	1	&	-1	&	1 && 1	&	-1	&	-1	&&	-1	&	1	&	1	&&	-1	&	1	&	-1	\\
		1	&	1	&	-1	&&	1	&	1	&	-1&& 1	&	-1	&	1	&&	1	&	-1	&	-1	&&	-1	&	1	&	1	\\
		1	&	1	&	1	&&	1	&	1	&	1 && 1	&	1	&	-1	&&	1	&	-1	&	-1	&&	-1	&	1	&	1	\\
		-1	&	-1	&	0	&&	-1	&	0	&	0 && 1	&	1	&	1	&&	1	&	-1	&	1	&&	1	&	-1	&	-1	\\
		-1	&	1	&	0	&&	-1	&	0	&	0 &&-1	&	-1	&	0	&&	1	&	-1	&	1	&&	1	&	-1	&	-1	\\
		1	&	-1	&	0	&&	-1	&	0	&	0 &&-1	&	1	&	0	&&	1	&	1	&	-1	&&	1	&	-1	&	1	\\
		1	&	1	&	0	&&	1	&	0	&	0 && 1	&	-1	&	0	& &	1	&	1	&	-1	&&	1	&	-1	&	1	\\
		-1	&	0	&	-1	&&	1	&	0	&	0 && 1	&	1	&	0	& &	1	&	1	&	1	&&	1	&	1	&	-1	\\
		-1	&	0	&	1	&&	1	&	0	&	0 &&-1	&	0	&	-1	& &	1	&	1	&	1	&&	1	&	1	&	-1	\\
		1	&	0	&	-1	&&	0	&	-1	&	0 &&-1	&	0	&	1	& &	-1	&	0	&	0	&&	1	&	1	&	1	\\
		1	&	0	&	1	&&	0	&	-1	&	0 && 1	&	0	&	-1	&&	-1	&	0	&	0	&&	1	&	1	&	1	\\
		0	&	-1	&	-1	&&	0	&	-1	&	0 && 1	&	0	&	1	&&	1	&	0	&	0	&&	-1	&	0	&	0	\\
		0	&	-1	&	1	&&	0	&	1	&	0 && 0	&	-1	&	-1	&&	1	&	0	&	0	&&	-1	&	0	&	0	\\
		0	&	1	&	-1	&&	0	&	1	&	0 && 0	&	-1	&	1	&&	0	&	-1	&	0	&&	1	&	0	&	0	\\
		0	&	1	&	1	&&	0	&	1	&	0 && 0	&	1	&	-1	&&	0	&	-1	&	0	&&	0	&	-1	&	0	\\
		0	&	0	&	0	&&	0	&	0	&	-1&& 0	&	1	&	1	&&	0	&	1	&	0	&&	0	&	-1	&	0	\\
		0	&	0	&	0	&&	0	&	0	&	-1&& 0	&	0	&	0	&&	0	&	1	&	0	&&	0	&	1	&	0	\\
		0	&	0	&	0	&&	0	&	0	&	-1&& 0	&	0	&	0	&&	0	&	0	&	-1	&&	0	&	0	&	-1	\\
		0	&	0	&	0	&&	0	&	0	&	1 && 0	&	0	&	0	&&	0	&	0	&	-1	&&	0	&	0	&	-1	\\
		0	&	0	&	0	&&	0	&	0	&	1 && 0	&	0	&	0	&&	0	&	0	&	1	&&	0	&	0	&	1	\\
		0	&	0	&	0	&&	0	&	0	&	1 && 0	&	0	&	0	&&	0	&	0	&	1	&&	0	&	0	&	1	\\
		\toprule 
	\end{tabular}
\end{table}
\begin{table}[h]
\scalefont{.8}
\caption{\label{tab:effEx1}Efficiencies of alternative designs for Example 1 ($n=26$, $q=3$, $p=10$ in cubic region)}
\vspace{0.2cm}
\centering
  % Esta tabela era grande demais para a página
%\begin{tabular}{lccccccccccccccccccccccccccccc}
\renewcommand{\tabcolsep}{0.1cm}
\begin{tabular}{cccrrrrrrrrrrrrrrrrrrrrrrrrrrrr}
\toprule
& & &\multicolumn{8}{c}{Efficiency}\\\cline{4-11}
Design&Criterion
 &\multicolumn{1}{c}{df(PE,~ LoF)$^{\dagger}$}&\multicolumn{1}{c}{$D_S$}&\multicolumn{1}{c}{$(DP)_S$}
      &\multicolumn{1}{c}{$A_S$}&\multicolumn{1}{c}{$(AP)_S$}&\multicolumn{1}{c}{$I$}
      &\multicolumn{1}{c}{$(IP)$}&\multicolumn{1}{c}{$I_D$}&\multicolumn{1}{c}{$(I_DP)$}   \\
    \midrule
1&{{$D_S$, $A_S$}}   &(~9,~~7)& 100.00&  86.77& 100.00&  95.50&  75.80&  72.32&  91.93&  87.00\\
2&{{$(DP)_S$}}  & (15,~~1)&  93.81& 100.00&  87.12&  93.72&  69.62&  74.82&  83.47&  88.98\\
3&{{$(AP)_S$}}  & (12,~~4)&  98.79&  97.45&  97.13& 100.00&  72.30&  74.36&  89.23&  91.02\\
4&{{$I$}}   & (~5,~11)&  90.71&  52.42&  87.71&  64.87& 100.00&  73.88&  99.87&  73.19\\
5&{{$(IP)$}}  & (12,~~4)&  79.79&  78.70&  72.80&  74.95&  97.23& 100.00&  87.47&  89.23\\
6&{{$I_D$}}  & (~5,~11)&   93.36&  53.96&  90.67&  67.06&  97.22&  71.83& 100.00&  73.28\\
7&{{$(I_DP)$}} & (12,~~4)&   95.29&  93.99&  92.11&  94.82&  92.00&  94.63&  98.03& 100.00\\
8&$\kappa_1=\kappa_7=0.5$  &(12,~4) &	 98.68&  97.34&  96.96&  99.82&  84.34&  86.74&  96.77&  98.71\\
9&{{$I_D,D_S,A_S$-$sym$}}  & (~5,~11)&   98.13&  56.71&  96.83&  71.62&  85.89&  63.46& 97.01&  71.09\\
\bottomrule 
\multicolumn{10}{l}{$ \dagger$df(PE,~LoF): degrees of freedom for pure error, degrees of freedom for lack of fit.}  \\
 \end{tabular}
\end{table}

Here we explore the prediction performances of some of the previously published designs and construct a few other alternatives based on estimation and prediction properties. The new designs are presented in Table \ref{tab:desEx1_1}. In Table \ref{tab:effEx1} we show the properties of the designs in terms of the usual single-valued criteria and the new criteria introduced in Section \ref{sec:DC}. 
Designs 1 to 3 were presented in \cite{gilmourtrinca2012}, design 9 is the best design \cite{borrotti2016} found for the properties $D_S$, $A_S$ and $I_D$, 
which they called the $I_D,D_S,A_S$-$symmetrical$ design. Designs from 4 to 8 are the new designs, the first four based on a single prediction property each ($I$, $(IP)$, $I_D$ and 
$(I_DP)$) and design 8 constructed by using a compound criterion with
$\kappa_1=\kappa_7=0.5$ in equation (\ref{eq:CP}), that is, giving equal priority for $(DP)_S$ and point predictions of difference of response.
%All nine designs are shown in Tables A and B in the Supplementary Material.

We note that, as the number of runs is not too small for the model specified, all designs allow for pure error degrees of freedom with designs 4, 6 and 9 
($I$, $I_D$ and $I_D,D_S,A_S$-$sym$) being the least attractive in this respect. Comparisons between designs 1 and 4 confirm the observation of \cite{jonesgoos2012} 
that the losses of $I$-optimum designs in terms of efficiencies for estimation, with respect to $D_S$ and $A_S$ criteria, are smaller than the 
losses of efficiencies in terms of prediction of $D_S$- or $A_S$-optimum designs. Similar lessons can be drawn when we compare designs 2 and 5 ($(DP)_S$- and $(IP)$-optimum 
designs) but now the differences are smaller. However, the results contradict the suggestion of Goos in the discussion of \cite{gilmourtrinca2012} that $I$-optimal designs usually have more replicates that $D$-optimal designs.

In general all designs based on a single property have low performance on at least one property except the
$(I_DP)$-optimum design which has a minimum efficiency of 92\%. However, in case we are interested in inferences for the parameters and predictions of differences in response, 
design 8 (obtained by the compound criterion, considering equal weights for $(DP)_S$ and $I_D$) has very high efficiencies for all properties. Surprisingly, 
design 8 outperforms design 9, the $I_D,D_S,A_S-sym$ multiple objective design from \cite{borrotti2016}, except for $I$ and $I_D$ properties, although the maximum 
difference between them in these two properties is only about 1.5\%. For properties like $(DP)_S$, $(AP)_S$,  $(I_DP)$ and $(IP)$ the advantage in using design
8 is overwhelming with efficiency gains of 40.63, 28.20, 27.62 and 23.28\%, respectively.   It is interesting to note that design 8 is very close to the
$(AP)_S$-optimum design (design 3) in terms of pure error and parameter estimation properties but it is considerably superior in terms of overall predictions. 
%These comparisons highlight the study of several designs properties in order 

\begin{figure}
\centering
\scalebox{0.49}[0.49]{\includegraphics{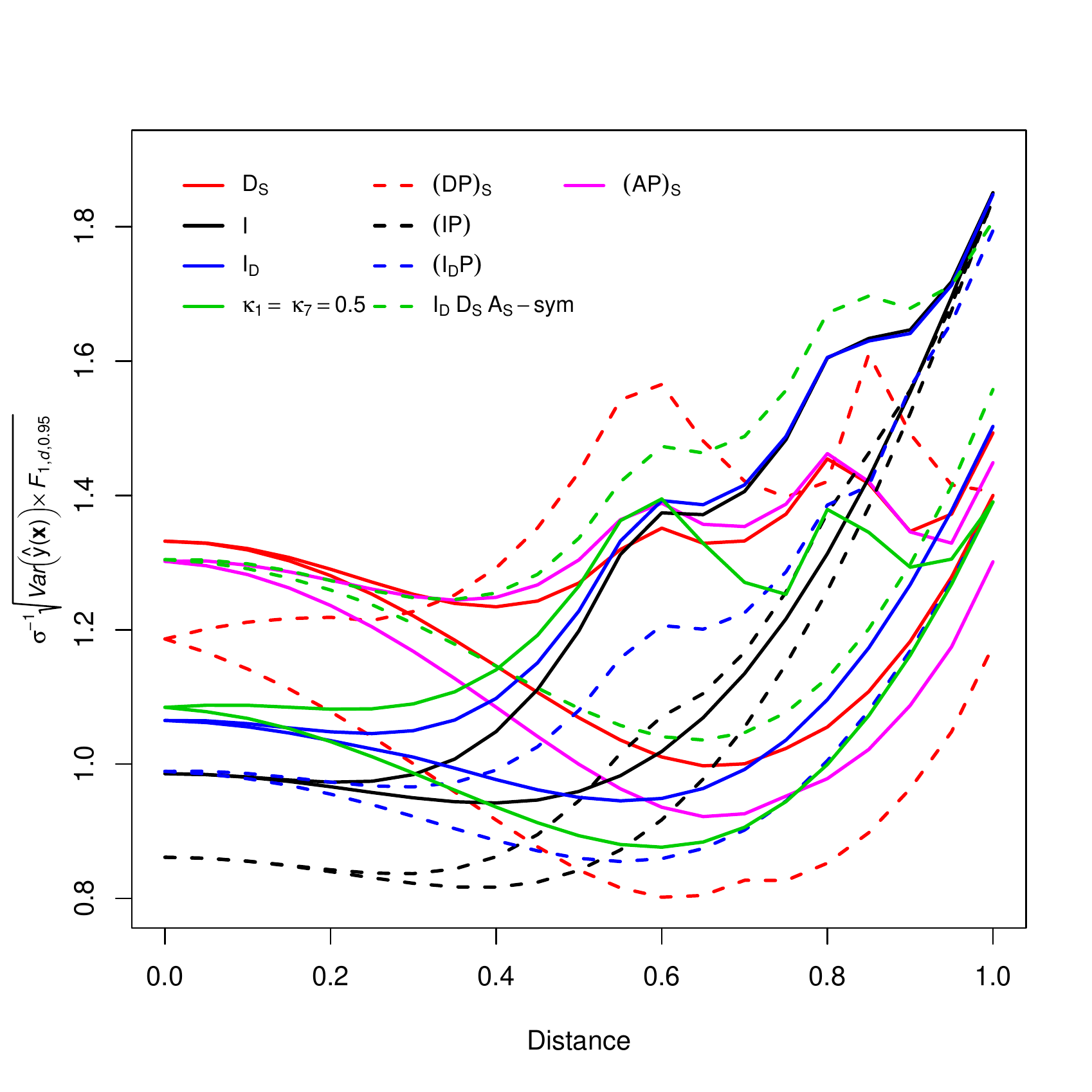}\includegraphics{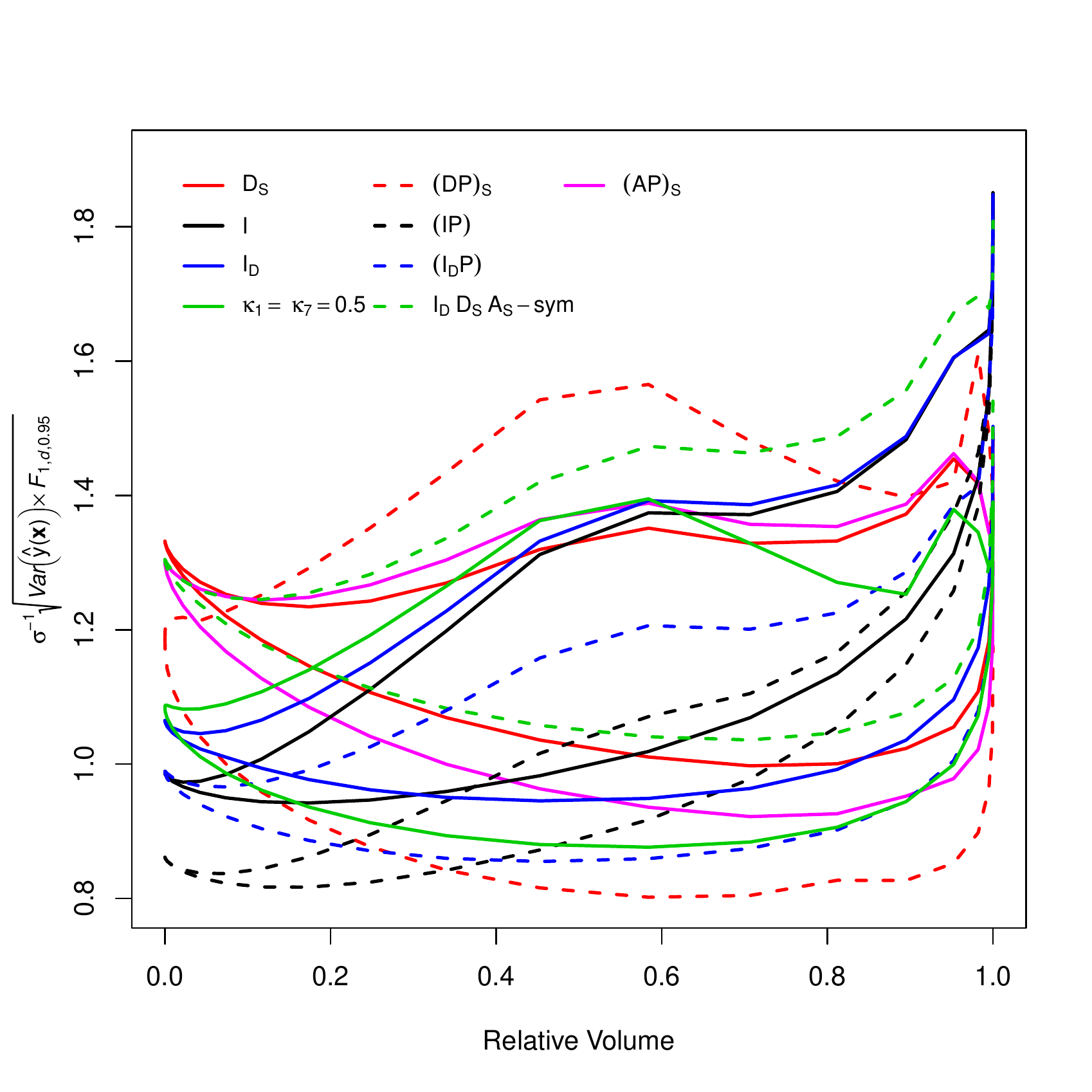}}
\caption{Standard error dispersion graphs (SEDG) of response predictions (interval), for designs in Example 1. Left: distance. Right: relative volume.}\label{graph:vdgpeEx1}
\end{figure}
\begin{figure}
	\scalebox{0.49}[0.49]{\includegraphics{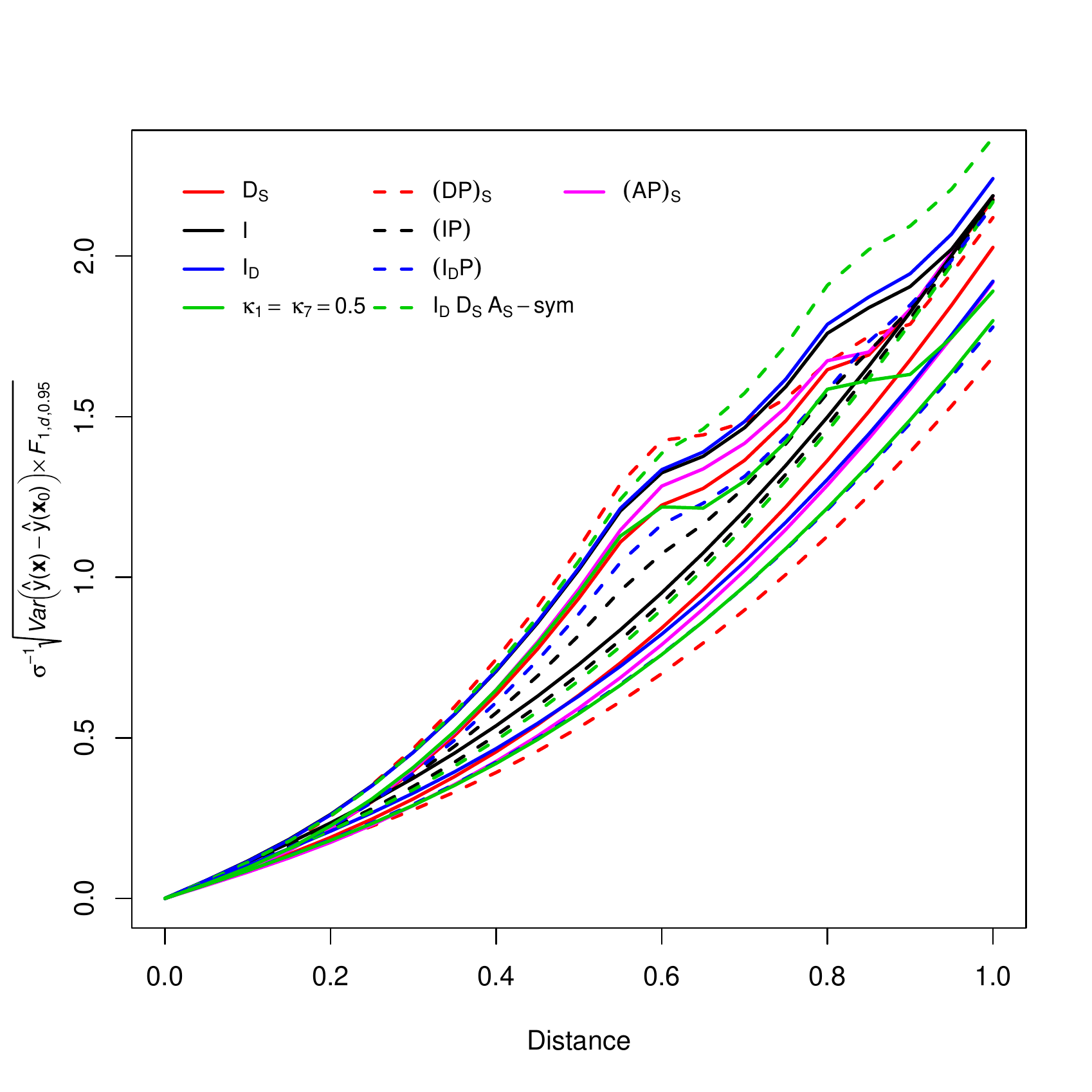}\includegraphics{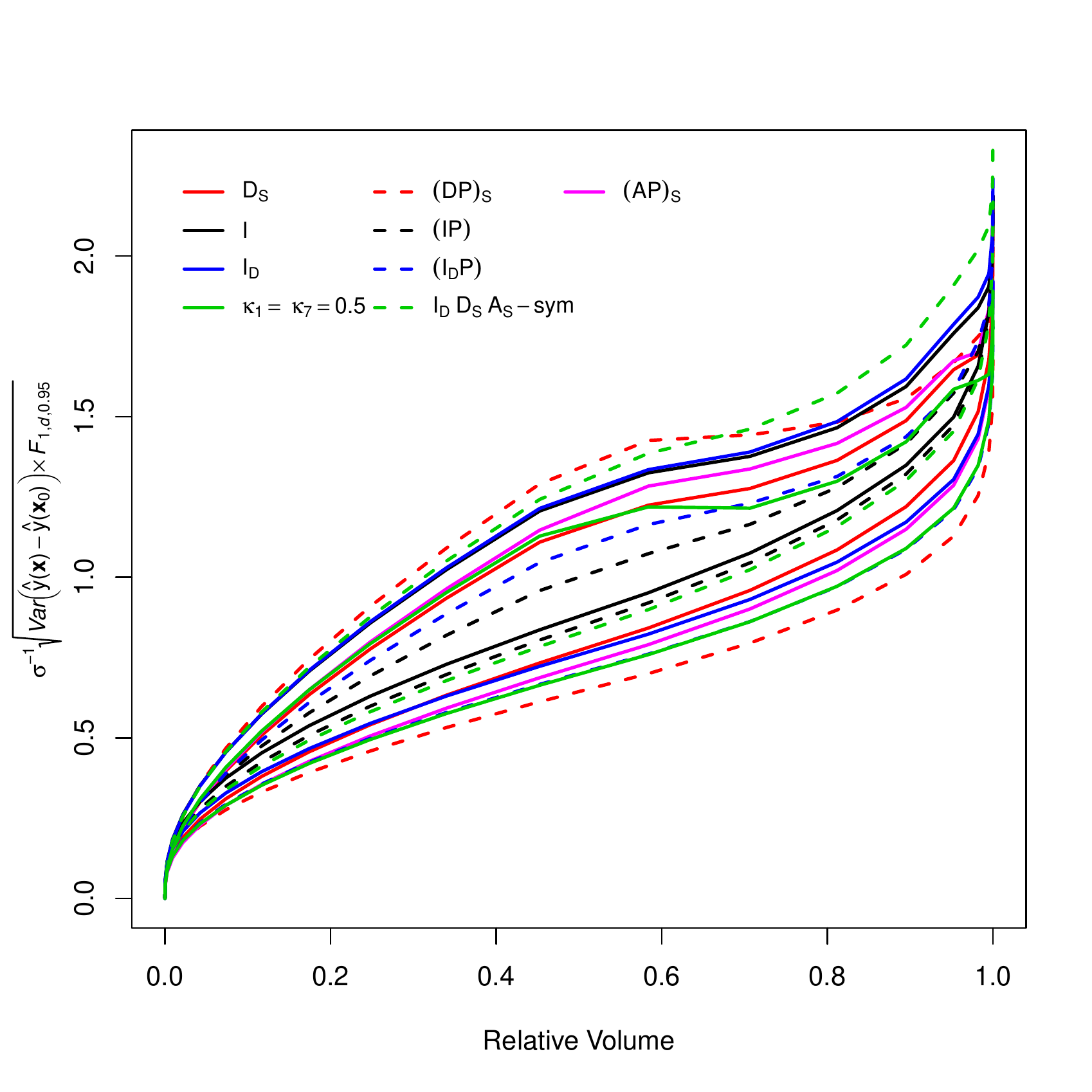}}
	\caption{Standard error dispersion graphs of differences (DSEDG) in response predictions (interval), for designs for Example 1.  Left: distance. Right: relative volume.}\label{graph:dvdgpeEx1}
\end{figure}

\begin{figure}
\centering
\scalebox{0.49}[0.49]{\includegraphics{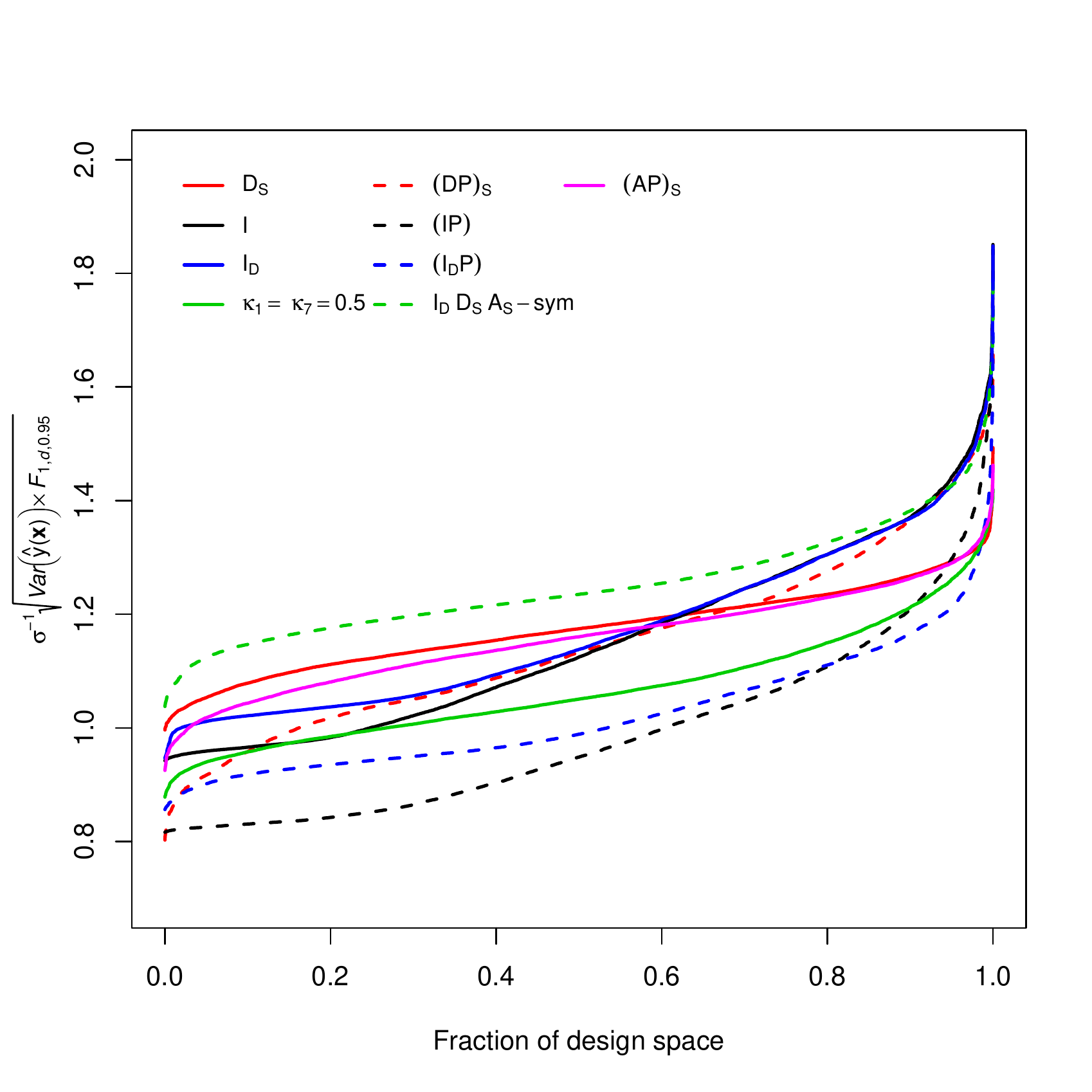}\includegraphics{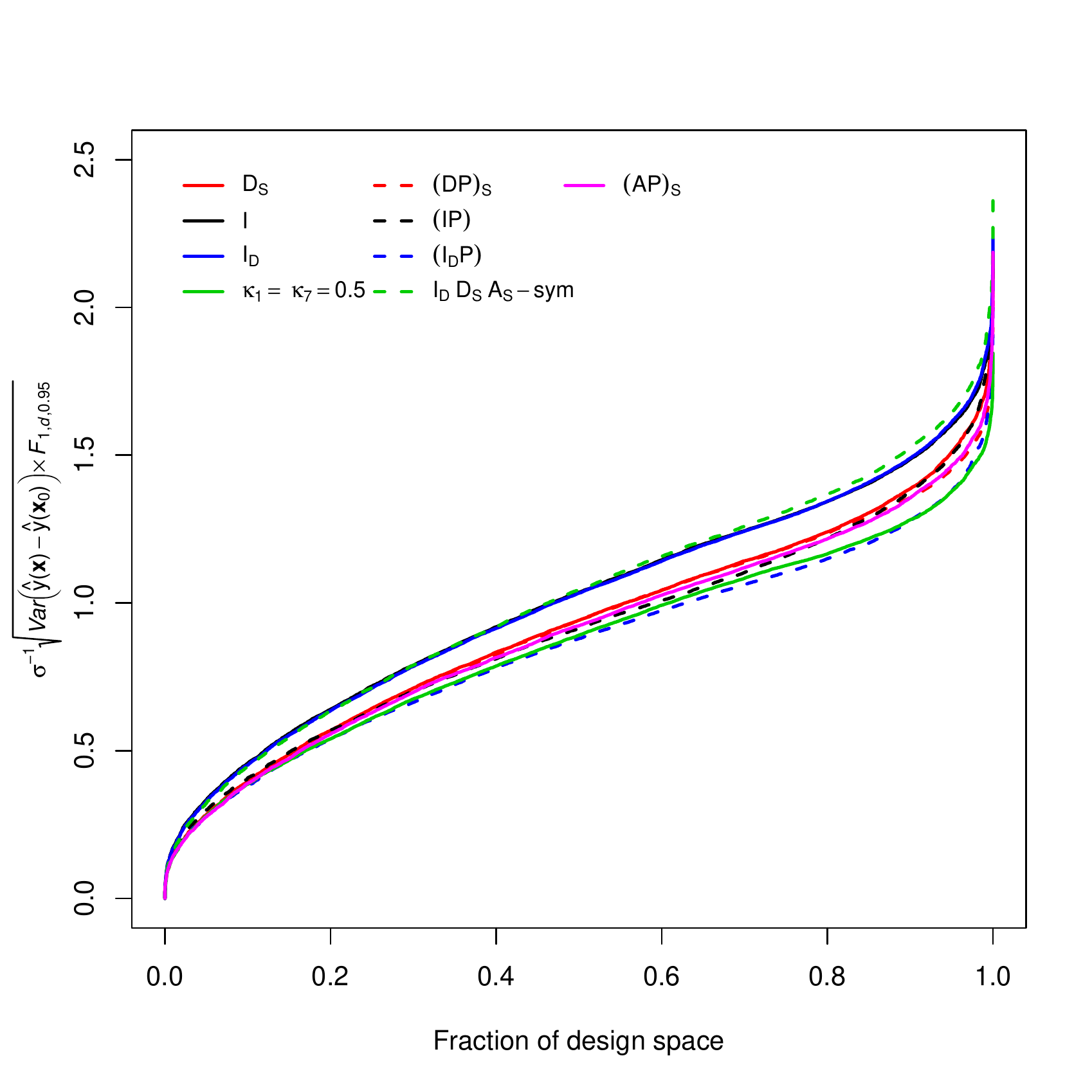}}
\caption{FDS plots, in terms of s.e., for designs in Example 1. Left: response interval predictions. Right: difference interval  predictions.}\label{graph:fdspeEx1}
\end{figure}

Figures \ref{graph:vdgpeEx1}-\ref{graph:fdspeEx1} (and Figures A-C in the Suppl.) show the prediction performances of the designs over the unit cube using standard error dispersion graphs (SEDGs). For the dispersion graphs (Figure A, left), the usual pattern is observed, i.e.\ the $(AP)_S$-, $D_S$- and $(DP)_S$-optimum designs have the highest s.e.\ at the center in order to control the precision in the corners. Several designs show two spikes around the relative distances of points in the cube face ($\approx 0.58$) and of points in the edges ($\approx 0.82$) with those of the $(DP)_S$-optimum design being most prominent. Note, however, that this design has the smallest minimum s.e.\ further from the center. In the other hand, the $I$-, $(IP)$- and $I_D$-optimum designs have the smallest s.e.'s in the middle but the s.e.'s are high for the portion away from the center. Our compound criterion design ($\kappa_1=\kappa_7=0.5$) does compromise and has similar performances to the  $I_D,D_S,A_S-sym$ design. Note however its superiority when interval prediction of responses is considered (Figure \ref{graph:vdgpeEx1}). The graph at the right hand-side of Figure \ref{graph:vdgpeEx1} presents the same information, but plotted against the relative volume contained within a radius, rather than its distance from the center. This variation of the plot seems more useful since it discriminates better between the designs.

The ordering of the designs in terms of response predictions is better summarized through the FDS graphs in Figures C (right) and \ref{graph:fdspeEx1} (right). It is interesting to note that the performance of the $(DP)_S$-optimum design is not as bad as suspected before. For interval predictions it outperforms design $I_D,D_S,A_S-sym$ in almost the whole region and outperforms the $D_S$-, $(AP)_S$- and $I_D$-optimum design in about $60\%$ of the region. Again, our compound criterion design compromises while the $(IP)$- and $(I_DP)$-optimum designs show the best performances overall.

The designs for Example 1 are quite homogeneous in terms of predictions of differences in the responses (see Figures \ref{graph:dvdgpeEx1},  \ref{graph:fdspeEx1} (right) and C (Supp) and the last two columns of Table \ref{tab:effEx1}). But we can still detect the superiority of our compound design and the $(I_DP)$-optimum design in the whole region.

\subsection{Example 2: $q=5$ factors in spherical region}

\begin{table}[hp]
	\scalefont{0.8}
	\caption{\label{tab:desEx2_1} Alternative designs for Example 2 ($n=30$, $q=5$, $p=21$ in spherical region)}
	\vspace{0.2cm}
	\centering
	\renewcommand{\arraystretch}{.7}
	\renewcommand{\tabcolsep}{0.1cm}
	\begin{tabular}{rrrrrccrrrrrccrrrrr}%		
		\toprule
		\multicolumn{19}{c}{Design}\\ \multicolumn{5}{c}{1}&&&\multicolumn{5}{c}{2}&&&\multicolumn{5}{c}{3} \\		
		\multicolumn{5}{c}{$D_S/I $}&&&\multicolumn{5}{c}{$(DP)_S$} &&&\multicolumn{5}{c}{$A_S$} \\
		
		\cmidrule(lr){1-5}\cmidrule(lr){8-12}\cmidrule(lr){15-19}
		$X_1$	&	$X_2$	&	$X_3$	&	$X_4$	&	$X_5$	&&&	$X_1$	&	$X_2$	&	$X_3$	&	$X_4$	&	$X_5$	&&&	$X_1$	&	$X_2$	&	$X_3$	&	$X_4$	&	$X_5$	\\ 
			\cmidrule(lr){1-5}\cmidrule(lr){8-12}\cmidrule(lr){15-19}
		-1.12	&	1.12	&	-1.12	&	0	&	-1.12	&&&	-1.29	&	-1.29	&	0	&	1.29	&	0	&&&	0	&	-1.12	&	1.12	&	-1.12	&	-1.12	\\
		-1.12	&	1.12	&	1.12	&	0	&	-1.12	&&&	-1.29	&	-1.29	&	0	&	-1.29	&	0	&&&	0	&	-1.12	&	1.12	&	1.12	&	-1.12	\\
		1.12	&	1.12	&	-1.12	&	0	&	-1.12	&&&	-1.29	&	1.29	&	0	&	-1.29	&	0	&&&	0	&	1.12	&	1.12	&	-1.12	&	-1.12	\\
		1.12	&	1.12	&	1.12	&	0	&	-1.12	&&&	-1.29	&	1.29	&	0	&	1.29	&	0	&&&	0	&	1.12	&	1.12	&	1.12	&	-1.12	\\
-1.29	&	1.29	&	0	&	0	&	1.29	&&&	-1.29	&	1.29	&	0	&	1.29	&	0	&&&	1.29	&	-1.29	&	-1.29	&	0	&	0	\\
1.29	&	1.29	&	0	&	0	&	1.29	&&&	1.29	&	-1.29	&	-1.29	&	0	&	0	&&&	1.29	&	-1.29	&	1.29	&	0	&	0	\\
-1.29	&	0	&	-1.29	&	1.29	&	0	&&&	1.29	&	1.29	&	-1.29	&	0	&	0	&&&	1.29	&	1.29	&	-1.29	&	0	&	0	\\
-1.29	&	0	&	1.29	&	1.29	&	0	&&&	1.29	&	1.29	&	-1.29	&	0	&	0	&&&	1.29	&	1.29	&	1.29	&	0	&	0	\\
1.29	&	0	&	-1.29	&	1.29	&	0	&&&	1.29	&	0	&	1.29	&	1.29	&	0	&&&	-1.29	&	-1.29	&	0	&	-1.29	&	0	\\
1.29	&	0	&	1.29	&	1.29	&	0	&&&	1.29	&	0	&	1.29	&	-1.29	&	0	&&&	-1.29	&	-1.29	&	0	&	1.29	&	0	\\
-1.29	&	0	&	0	&	-1.29	&	-1.29	&&&	1.29	&	0	&	1.29	&	1.29	&	0	&&&	-1.29	&	1.29	&	0	&	-1.29	&	0	\\
-1.29	&	0	&	0	&	-1.29	&	1.29	&&&	0	&	-1.29	&	1.29	&	0	&	-1.29	&&&	-1.29	&	1.29	&	0	&	1.29	&	0	\\
1.29	&	0	&	0	&	-1.29	&	-1.29	&&&	0	&	-1.29	&	1.29	&	0	&	1.29	&&&	-1.29	&	0	&	-1.29	&	0	&	-1.29	\\
1.29	&	0	&	0	&	-1.29	&	1.29	&&&	0	&	-1.29	&	1.29	&	0	&	1.29	&&&	-1.29	&	0	&	-1.29	&	0	&	1.29	\\
0	&	-1.29	&	-1.29	&	-1.29	&	0	&&&	0	&	1.29	&	1.29	&	0	&	-1.29	&&&	-1.29	&	0	&	1.29	&	0	&	-1.29	\\
0	&	-1.29	&	1.29	&	-1.29	&	0	&&&	0	&	1.29	&	1.29	&	0	&	1.29	&&&	-1.29	&	0	&	1.29	&	0	&	1.29	\\
0	&	1.29	&	-1.29	&	-1.29	&	0	&&&	0	&	1.29	&	1.29	&	0	&	1.29	&&&	1.29	&	0	&	0	&	-1.29	&	-1.29	\\
0	&	1.29	&	1.29	&	-1.29	&	0	&&&	0	&	0	&	-1.29	&	-1.29	&	-1.29	&&&	1.29	&	0	&	0	&	-1.29	&	1.29	\\
0	&	-1.29	&	-1.29	&	0	&	-1.29	&&&	0	&	0	&	-1.29	&	-1.29	&	1.29	&&&	1.29	&	0	&	0	&	1.29	&	-1.29	\\
0	&	-1.29	&	1.29	&	0	&	-1.29	&&&	0	&	0	&	-1.29	&	-1.29	&	-1.29	&&&	1.29	&	0	&	0	&	1.29	&	1.29	\\
-1.58	&	-1.58	&	0	&	0	&	0	&&&	0	&	0	&	-1.29	&	1.29	&	1.29	&&&	0	&	-1.29	&	-1.29	&	0	&	-1.29	\\
1.58	&	-1.58	&	0	&	0	&	0	&&&	0	&	0	&	-1.29	&	1.29	&	-1.29	&&&	0	&	1.29	&	-1.29	&	0	&	-1.29	\\
0	&	-1.58	&	0	&	1.58	&	0	&&&	0	&	0	&	-1.29	&	-1.29	&	1.29	&&&	0	&	0	&	1.29	&	-1.29	&	1.29	\\
0	&	1.58	&	0	&	1.58	&	0	&&&	-1.58	&	0	&	-1.58	&	0	&	0	&&&	0	&	0	&	1.29	&	1.29	&	1.29	\\
0	&	-1.58	&	0	&	0	&	1.58	&&&	-1.58	&	0	&	1.58	&	0	&	0	&&&	0	&	-1.58	&	0	&	0	&	1.58	\\
0	&	0	&	-1.58	&	0	&	1.58	&&&	-1.58	&	0	&	1.58	&	0	&	0	&&&	0	&	1.58	&	0	&	0	&	1.58	\\
0	&	0	&	1.58	&	0	&	1.58	&&&	1.58	&	0	&	0	&	0	&	-1.58	&&&	0	&	0	&	-1.58	&	-1.58	&	0	\\
0	&	0	&	0	&	1.58	&	-1.58	&&&	1.58	&	0	&	0	&	0	&	1.58	&&&	0	&	0	&	-1.58	&	1.58	&	0	\\
0	&	0	&	0	&	1.58	&	1.58	&&&	1.58	&	0	&	0	&	0	&	1.58	&&&	0	&	0	&	0	&	0	&	0	\\
0	&	0	&	0	&	0	&	0	&&&	0	&	0	&	0	&	0	&	0	&&&	0	&	0	&	0	&	0	&	0	\\
\toprule 
\end{tabular}
\end{table}

\begin{table}[hp]
	\scalefont{0.8}
	\caption{\label{tab:desEx2_2} Alternative designs for Example 2 ($n=30$, $q=5$, $p=21$ in spherical region) }
	\vspace{0.2cm}
\centering
\renewcommand{\arraystretch}{.7}
\renewcommand{\tabcolsep}{0.1cm}
\begin{tabular}{rrrrrccrrrrrccrrrrr}%
		\toprule
		\multicolumn{19}{c}{Design}\\ \multicolumn{5}{c}{4}&&&\multicolumn{5}{c}{5}&&&\multicolumn{5}{c}{7} \\		
		\multicolumn{5}{c}{$(AP)_S$}&&&\multicolumn{5}{c}{$(IP)$} &&&\multicolumn{5}{c}{$(I_DP)$} \\	
\cmidrule(lr){1-5}\cmidrule(lr){8-12}\cmidrule(lr){15-19}
$X_1$	&	$X_2$	&	$X_3$	&	$X_4$	&	$X_5$	&&&	$X_1$	&	$X_2$	&	$X_3$	&	$X_4$	&	$X_5$	&&&	$X_1$	&	$X_2$	&	$X_3$	&	$X_4$	&	$X_5$\\
		\cmidrule(lr){1-5}\cmidrule(lr){8-12}\cmidrule(lr){15-19}
-1&-1&-1&-1&-1&&&-1&-1&-1&-1&1&&&-1&-1&-1&-1&1\\
-1&-1&-1&1&1&&&-1&-1&-1&-1&1&&&-1&-1&-1&1&-1\\
-1&-1&1&-1&1&&&-1&-1&-1&1&-1&&&-1&-1&-1&1&-1\\
-1&-1&1&-1&1&&&-1&-1&1&-1&-1&&&-1&-1&1&-1&-1\\
-1&-1&1&1&-1&&&-1&-1&1&-1&-1&&&-1&-1&1&1&1\\
-1&1&-1&-1&1&&&-1&-1&1&1&1&&&-1&1&-1&-1&-1\\
-1&1&-1&1&-1&&&-1&1&-1&1&1&&&-1&1&-1&1&1\\
-1&1&1&-1&-1&&&-1&1&1&-1&1&&&-1&1&1&-1&1\\
-1&1&1&1&1&&&-1&1&1&1&-1&&&-1&1&1&1&-1\\
1&-1&-1&-1&1&&&1&-1&-1&1&1&&&-1&1&1&1&-1\\
1&-1&-1&1&-1&&&1&-1&1&-1&1&&&1&-1&-1&-1&-1\\
1&-1&-1&1&-1&&&1&-1&1&-1&1&&&1&-1&-1&1&1\\
1&-1&1&-1&-1&&&1&-1&1&1&-1&&&1&-1&1&-1&1\\
1&-1&1&1&1&&&1&1&-1&-1&1&&&1&-1&1&-1&1\\
1&1&-1&1&1&&&1&1&-1&1&-1&&&1&-1&1&1&-1\\
1&1&-1&1&1&&&1&1&-1&1&-1&&&1&1&-1&-1&1\\
1&1&1&-1&1&&&1&1&1&-1&-1&&&1&1&1&-1&-1\\
1&1&1&1&-1&&&1&1&1&-1&-1&&&1&1&1&1&1\\
1&1&1&1&-1&&&1&1&1&1&1&&&0&1.12&-1.12&1.12&-1.12\\
1.12&1.12&-1.12&0&-1.12&&&1&1&1&1&1&&&0&1.12&-1.12&1.12&-1.12\\
1.12&1.12&-1.12&0&-1.12&&&-1.12&1.12&-1.12&0&-1.12&&&2.24&0&0&0&0\\
2.24&0&0&0&0&&&-1.12&1.12&-1.12&0&-1.12&&&0&-2.24&0&0&0\\
0&-2.24&0&0&0&&&0&-1.12&-1.12&-1.12&-1.12&&&0&0&2.24&0&0\\
0&0&2.24&0&0&&&0&-1.12&-1.12&-1.12&-1.12&&&0&0&0&2.24&0\\
0&0&0&-2.24&0&&&2.24&0&0&0&0&&&0&0&0&0&2.24\\
0&0&0&0&2.24&&&0&2.24&0&0&0&&&0&0&0&0&0\\
0&0&0&0&0&&&0&0&-2.24&0&0&&&0&0&0&0&0\\
0&0&0&0&0&&&0&0&0&-2.24&0&&&0&0&0&0&0\\
0&0&0&0&0&&&0&0&0&0&2.24&&&0&0&0&0&0\\
0&0&0&0&0&&&0&0&0&0&0&&&0&0&0&0&0\\
		\toprule 
	\end{tabular}
\end{table}

\begin{table}[hp]
	\scalefont{.8}
	\caption{\label{tab:desEx2_3} Alternative designs for Example 2 ($n=30$, $q=5$, $p=21$ in spherical region)}
	\vspace{0.2cm}
	\centering
	\renewcommand{\arraystretch}{.7}
	\renewcommand{\tabcolsep}{0.1cm}
\begin{tabular}{rrrrrccrrrrrccrrrrr}%	
			\toprule
		\multicolumn{19}{c}{Design}\\ \multicolumn{5}{c}{8}&&&\multicolumn{5}{c}{9}&&&\multicolumn{5}{c}{10} \\		
		\multicolumn{5}{c}{$\kappa_1=.3;~\kappa_7=.7$}&&&\multicolumn{5}{c}{$\kappa_1=.1;~\kappa_7=.9$} &&&\multicolumn{5}{c}{$\kappa_0=.9;~\kappa_8=.1$
		} \\
	\cmidrule(lr){1-5}\cmidrule(lr){8-12}\cmidrule(lr){15-19}
		$X_1$	&	$X_2$	&	$X_3$	&	$X_4$	&	$X_5$	&&&	$X_1$	&	$X_2$	&	$X_3$	&	$X_4$	&	$X_5$	&&&	$X_1$	&	$X_2$	&	$X_3$	&	$X_4$	&	$X_5$	\\	\cmidrule(lr){1-5}\cmidrule(lr){8-12}\cmidrule(lr){15-19}
1.12&-1.12&1.12&1.12&0&&&-1&-1&-1&1&-1&&&-1.29&-1.29&-1.29&0&0\\
1.12&1.12&1.12&1.12&0&&&-1&-1&1&1&1&&&-1.29&-1.29&1.29&0&0\\
1.12&-1.12&-1.12&0&1.12&&&-1&1&-1&1&1&&&-1.29&1.29&-1.29&0&0\\
1.12&1.12&-1.12&0&1.12&&&-1&1&1&1&-1&&&-1.29&1.29&1.29&0&0\\
-1.29&-1.29&-1.29&0&0&&&-1&1&1&-1&1&&&1.29&-1.29&0&0&-1.29\\
-1.29&-1.29&1.29&0&0&&&1&-1&-1&-1&-1&&&1.29&-1.29&0&0&-1.29\\
-1.29&1.29&-1.29&0&0&&&1&-1&1&-1&1&&&1.29&-1.29&0&0&1.29\\
-1.29&1.29&1.29&0&0&&&1&1&-1&-1&1&&&1.29&-1.29&0&0&1.29\\
-1.29&0&0&-1.29&-1.29&&&1&1&1&1&1&&&1.29&1.29&0&0&-1.29\\
-1.29&0&0&-1.29&1.29&&&1&1&1&-1&-1&&&1.29&1.29&0&0&1.29\\
-1.29&0&0&1.29&-1.29&&&-1.29&-1.29&0&-1.29&0&&&-1.29&0&0&-1.29&1.29\\
-1.29&0&0&1.29&1.29&&&1.29&-1.29&0&1.29&0&&&-1.29&0&0&1.29&-1.29\\
-1.29&0&0&1.29&1.29&&&-1.29&0&-1.29&-1.29&0&&&-1.29&0&0&1.29&1.29\\
1.29&0&-1.29&0&-1.29&&&1.29&0&-1.29&1.29&0&&&-1.29&0&0&-1.29&-1.29\\
1.29&0&1.29&-1.29&0&&&-1.29&0&0&-1.29&-1.29&&&1.29&0&-1.29&-1.29&0\\
0&-1.58&0&-1.58&0&&&1.29&0&0&1.29&-1.29&&&1.29&0&-1.29&1.29&0\\
0&1.58&0&-1.58&0&&&0&-1.29&-1.29&0&1.29&&&1.29&0&1.29&-1.29&0\\
0&-1.58&0&0&-1.58&&&0&-1.29&1.29&0&-1.29&&&1.29&0&1.29&-1.29&0\\
0&1.58&0&0&-1.58&&&0&1.29&-1.29&0&-1.29&&&1.29&0&1.29&1.29&0\\
0&0&-1.58&-1.58&0&&&-2.24&0&0&0&0&&&0&-1.58&0&-1.58&0\\
0&0&-1.58&1.58&0&&&0&2.24&0&0&0&&&0&-1.58&0&1.58&0\\
0&0&-1.58&1.58&0&&&0&0&2.24&0&0&&&0&1.58&0&-1.58&0\\
0&0&1.58&0&-1.58&&&0&0&0&2.24&0&&&0&1.58&0&1.58&0\\
0&0&1.58&0&1.58&&&0&0&0&0&2.24&&&0&0&-1.58&0&-1.58\\
0&0&1.58&0&1.58&&&0&0&0&0&0&&&0&0&-1.58&0&1.58\\
0&0&0&0&0&&&0&0&0&0&0&&&0&0&1.58&0&-1.58\\
0&0&0&0&0&&&0&0&0&0&0&&&0&0&1.58&0&1.58\\
0&0&0&0&0&&&0&0&0&0&0&&&0&0&0&0&0\\
0&0&0&0&0&&&0&0&0&0&0&&&0&0&0&0&0\\
0&0&0&0&0&&&0&0&0&0&0&&&0&0&0&0&0\\
		\toprule 
	\end{tabular}
\end{table}

\cite{jang2012} compared a few classical designs (CCD, Box-Behnken design) for five factors in a spherical region considering several run sizes. Here we constructed several optimum designs for $n=30$ and the second order model ($p=21$) and we compare them with the resolution-V half fraction CCD ($\alpha = \sqrt{5} \approx 2.236$) with four center runs. The designs are shown in Tables \ref{tab:desEx2_1}, \ref{tab:desEx2_2} and \ref{tab:desEx2_3}. Interestingly we found that the $I_D$-optimum design is the resolution-V CCD, which is very unusual for an optimum design chosen from such a large candidate set. We found other equivalences among designs, for example the $D_S$-optimum design is also $I$-optimum, although, since we are using heuristics, we have no absolute guarantee that the true optimum designs for these criteria are equivalent or unique. Design 11 is also similar to a CCD except that it includes four factorial points duplicated (see Table \ref{d11}), the center point is replicated four times and includes the axial pair for only one factor ($X_3$), while for the other factors it includes only one axial point. 
\begin{table}
		\scalefont{0.8}
	\caption{\label{d11} Points from the $2^5$ that are duplicated in Design 11 (Table \ref{tab:effEx2})}
	\vspace{0.2cm}
	\centering
	\renewcommand{\arraystretch}{.7}
	\renewcommand{\tabcolsep}{0.1cm}
	\begin{tabular}{rrrrr}%		
		\toprule
			$X_1$	&	$X_2$	&	$X_3$	&	$X_4$	&	$X_5$	\\
			\toprule
	-1	&	-1	&	1	&	1	&	1\\
	-1	&	1	&	-1	&	1	&	1\\
	-1	&	1	&	1	&	1	&	-1\\
	1	&	-1	&	1	&	-1	&	1\\
				\toprule
\end{tabular}	
\end{table} 

\begin{table}[ht]
	\scalefont{.75}
	\caption{\label{tab:effEx2}Efficiencies of alternative designs for Example 2 ($n=30$, $q=5$, $p=21$ in spherical region)}
	\vspace{0.2cm}
	\centering
	\renewcommand{\arraystretch}{.7}
	\renewcommand{\tabcolsep}{0.03cm}
	\begin{tabular}{cccrrrrrrrr}
		\toprule
		& & &\multicolumn{8}{c}{Efficiency}\\\cline{4-11}
		Design&Criterion
		&\multicolumn{1}{c}{df(PE,~ LoF)$^{\dagger}$}&\multicolumn{1}{c}{$D_S$}&\multicolumn{1}{c}{$(DP)_S$}
		&\multicolumn{1}{c}{$A_S$}&\multicolumn{1}{c}{$(AP)_S$}&\multicolumn{1}{c}{$I$}
		&\multicolumn{1}{c}{$(IP)$}&\multicolumn{1}{c}{$I_D$}&\multicolumn{1}{c}{$(I_DP)$} \\
		\midrule
		1&{{$D_S$, $I$}}   & (0,~9)&   100.00&   0.00&  94.02&   0.00& 100.00&   0.00&  60.31&   0.00\\
		2&{{$(DP)_S$}}  & (9,~0)&  86.30 & 100.00&  74.33&  90.36&  74.73 & 97.81 &  52.80&  65.56\\
		3&{{ $A_S$}}   &(1,~8)&   98.16  & 1.35&   100.00&    3.85&   92.86&     3.85&    81.20&    3.10\\
		4&{{$(AP)_S$}} &(8,~1) & 87.39 & 94.39  &85.48&100.00  &74.34&93.64&844.84 &98.28\\
		5&{{$(IP)$}}  & (8,~1)&  88.84&  95.95&  79.04&  92.47&  79.39& 100.00&  54.37&  62.99\\
		6&{{CCD, $I_D$}}  & (3,~6)&    96.96&  38.09&  95.25&  58.51&  91.82&  60.73& 100.00&  60.82\\
		7&{$(I_DP)$}&(8,~1)&85.37&92.20&83.63&97.83&72.21&90.95&86.32&100.00\\
		8&$\kappa_1=0.3;~\kappa_7=0.7$&(7,~2)&85.74&84.69&82.89&92.22&73.35&87.87&87.46&96.35\\
		9&$\kappa_1=0.1;~\kappa_7=0.9$&(5,~4)&86.71&64.73&85.61&80.60&76.58&77.62&93.34&87.02\\
		10&$\kappa_0=0.9;~\kappa_8=0.1$&(5,~4)&93.49&  69.79 & 91.88&  86.50&  84.56 & 85.72&  87.32&  81.40\\
		11&$\kappa_0=\kappa_1=.2;~\kappa_3=\kappa_6=.3$&(7,~2)&90.35&89.25&88.94&98.95&78.50&94.03&88.57&  97.58\\		
		\bottomrule 
		\multicolumn{10}{l}{$ \dagger$df(PE,~LoF): degrees of freedom for pure error, degrees of freedom for lack of fit.}  \\
	\end{tabular}
\end{table}

The efficiencies of several designs are shown in Table \ref{tab:effEx2}. The optimum designs from the usual criteria do not allow pure error estimation ($D_S/I$) or provide very few treatment replications ($A_S$ and CCD/$I_D$) and thus, efficiencies of these designs with respect to modified criteria are zero or small. We note that designs $(AP)_S$, $(I_DP)$ are similar and have reasonably high efficiencies generally, providing 8 degrees of freedom for error estimation but only one spare degree of freedom to add a higher order term in the model in case experimental results show lack of fit of the quadratic model. Design 11 behaves similarly but has the advantage of allowing two degrees of freedom for lack of fit. We tried many weight patterns for this example to obtain compromise designs but many returned designs equivalent to some of the single property criteria and so, we present results for only four of them, designs 8, 9, 10 and 11. From these we see that designs 9 and 10 balance better the degrees of freedom between pure error and lack of fit. Design 10, which focuses on parameter estimation through the $D_S$ criterion and interval estimation of differences in response, has reasonably high efficiencies overall.

\begin{figure}[H]
\centering
		\scalebox{0.49}[0.49]{\includegraphics{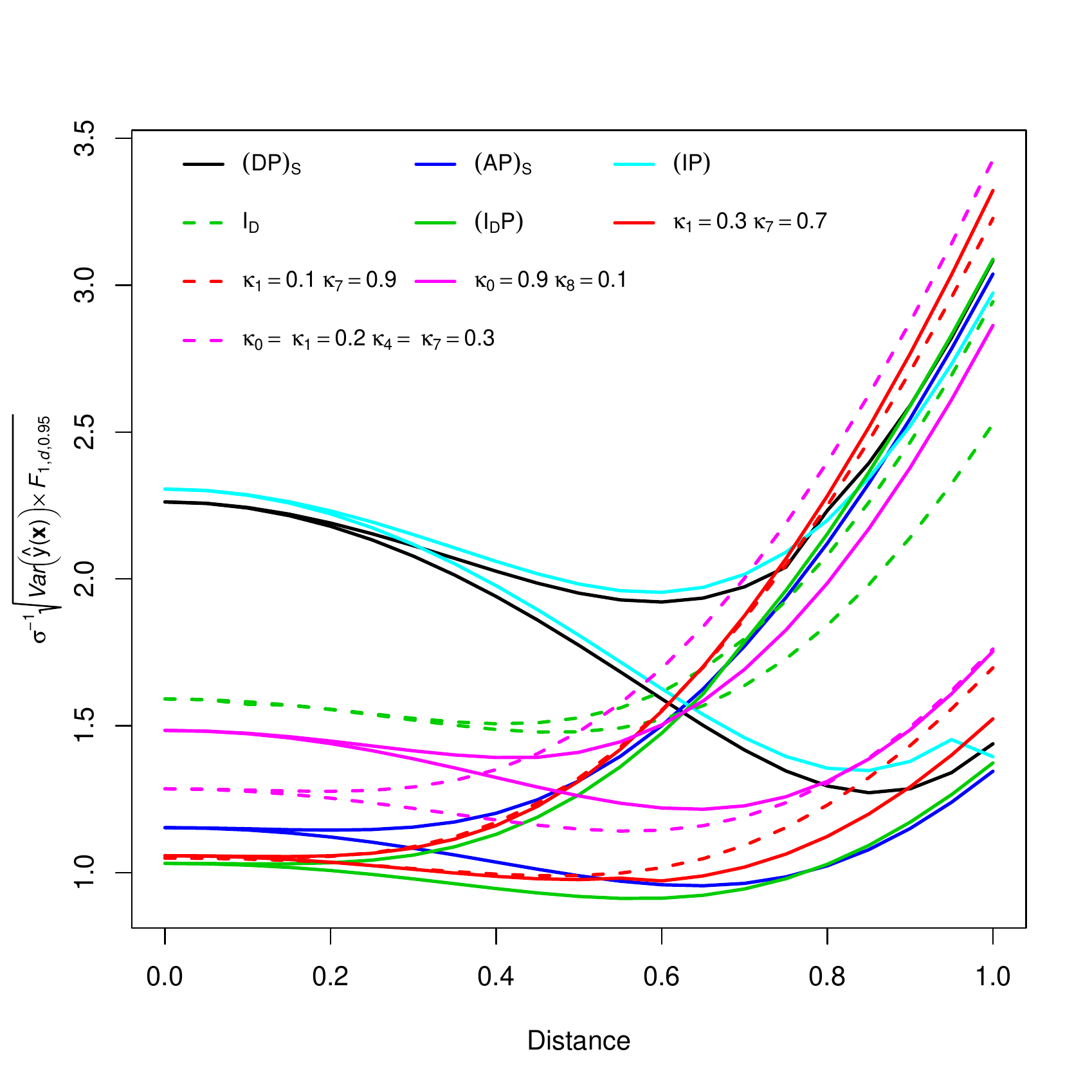}\includegraphics{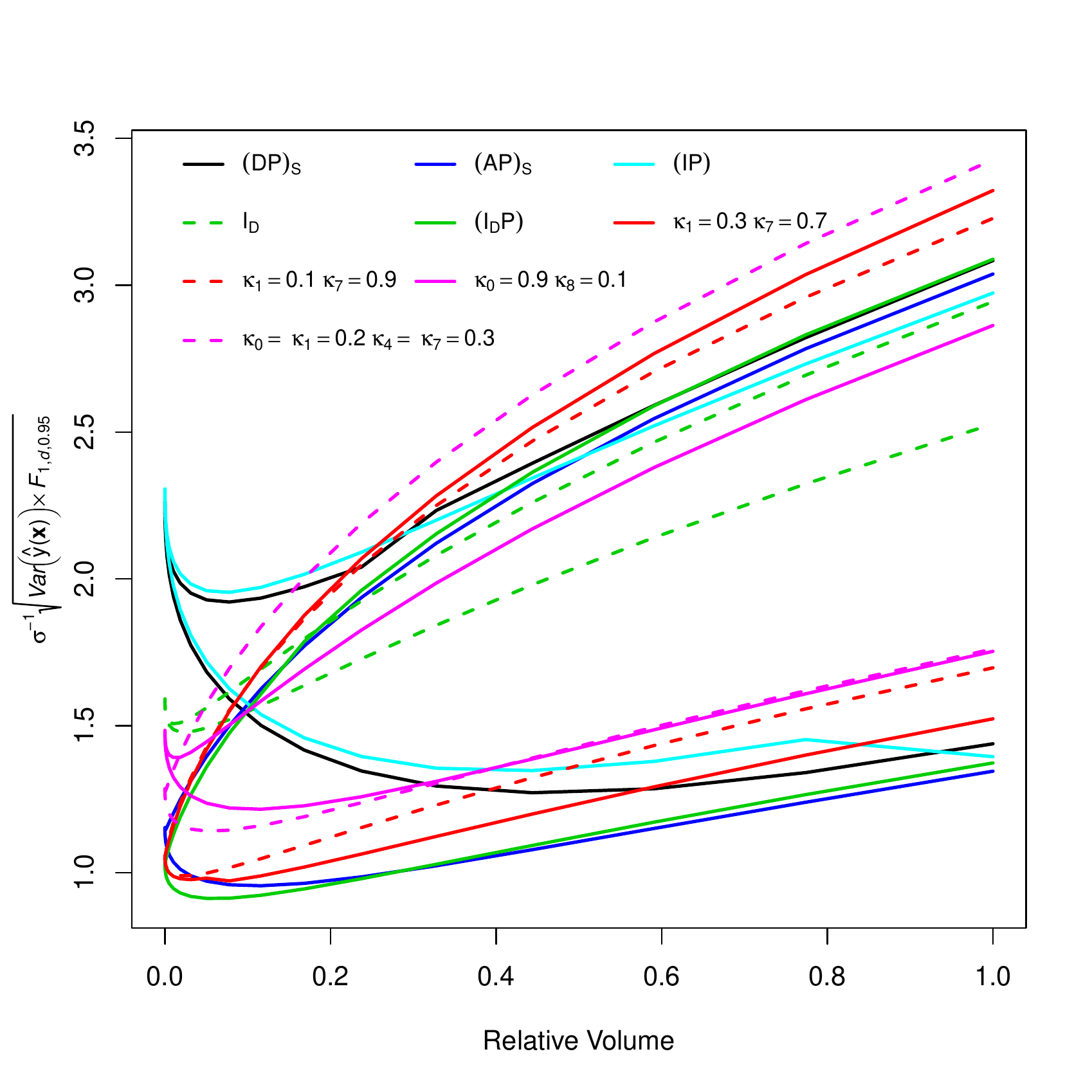}}
		\caption{SEDGs (interval) for designs in Example 2. Left: distance. Right: relative volume.}\label{graph:vdgpeEx2}
\end{figure}
\begin{figure}
		\scalebox{0.49}[0.49]{\includegraphics{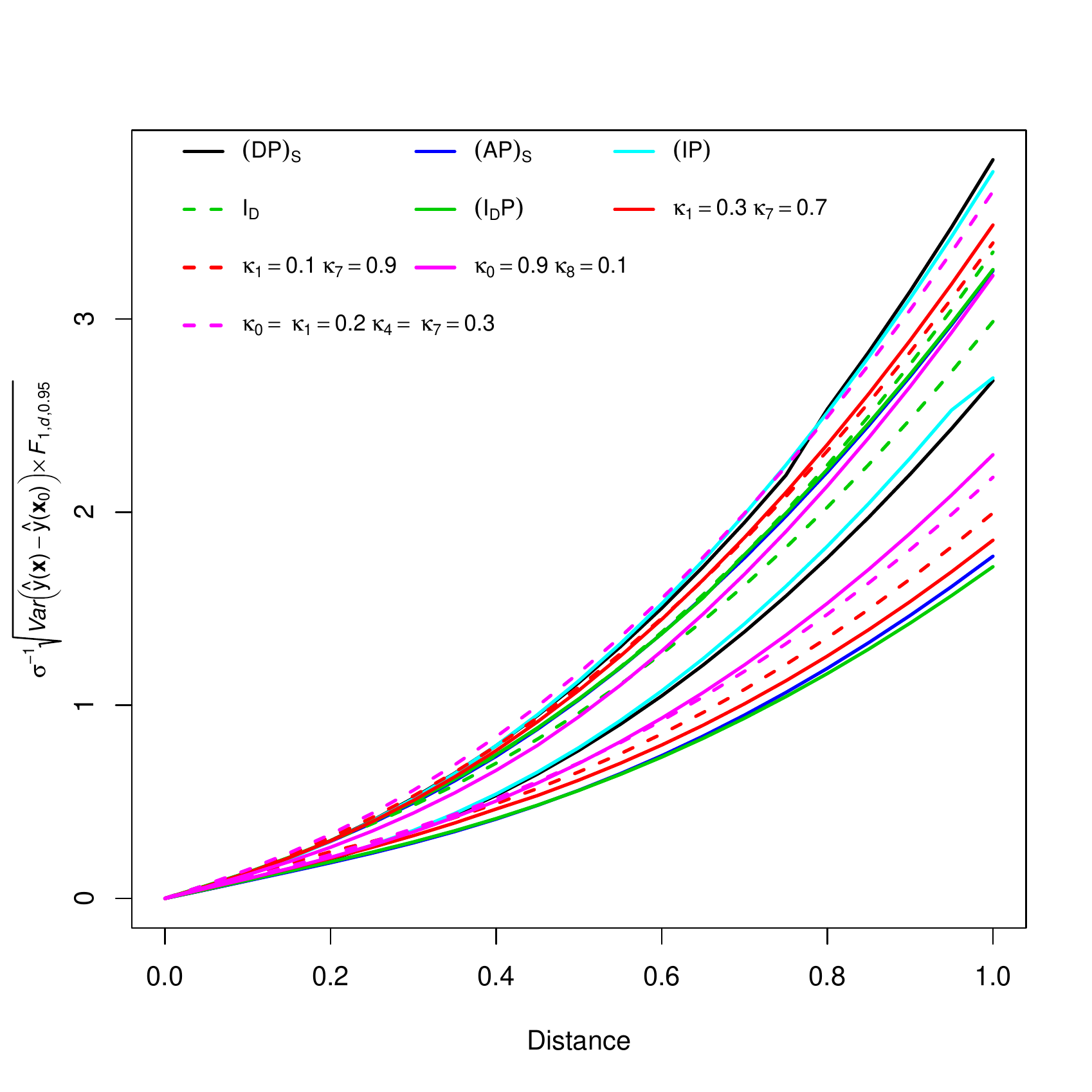}\includegraphics{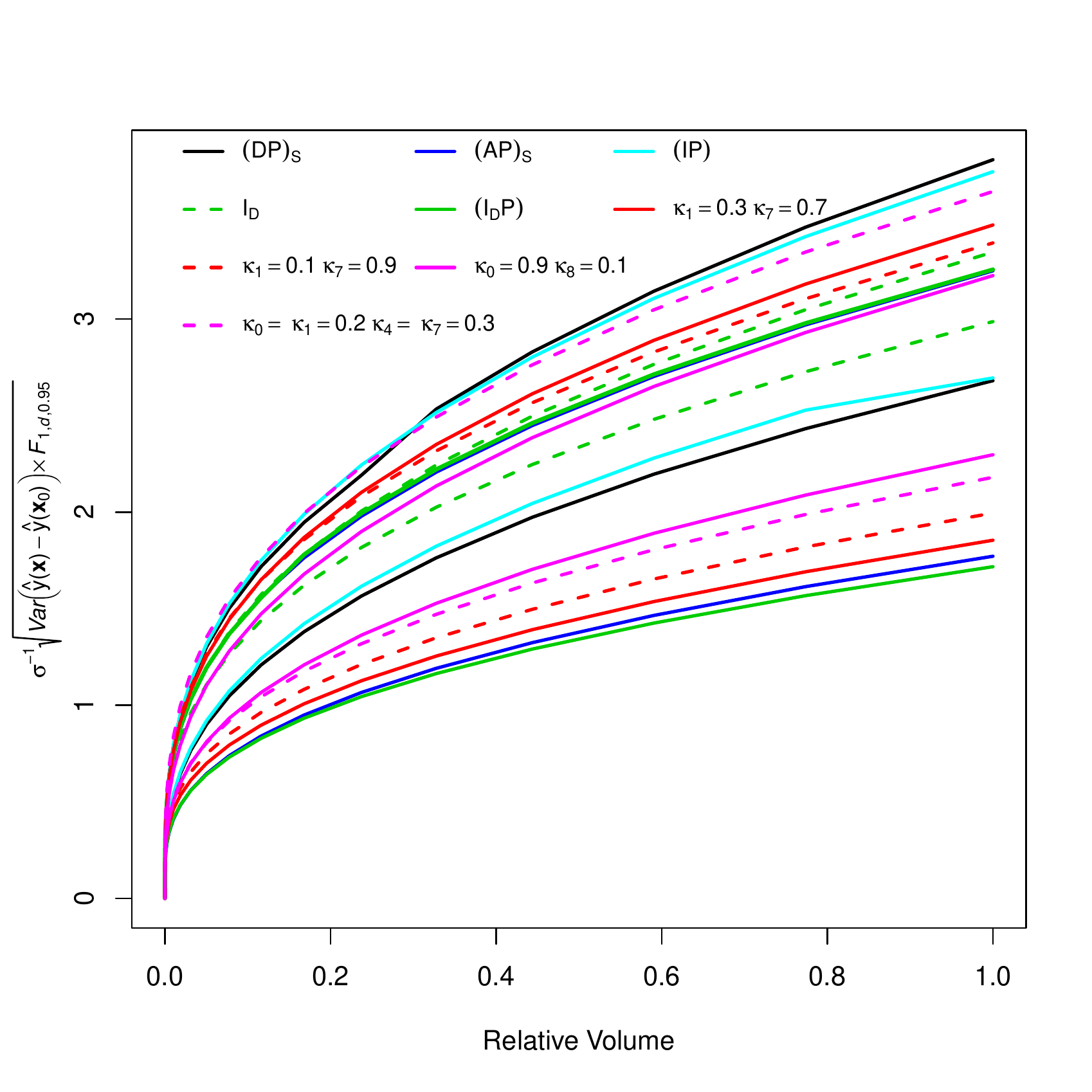}}
		\caption{DSEDGs (interval) for designs in Example 2. Left: distance. Right: relative volume.}\label{graph:dvdgpeEx2}
\end{figure}  

\begin{figure}
	\begin{center}
		\scalebox{0.49}[0.49]{\includegraphics{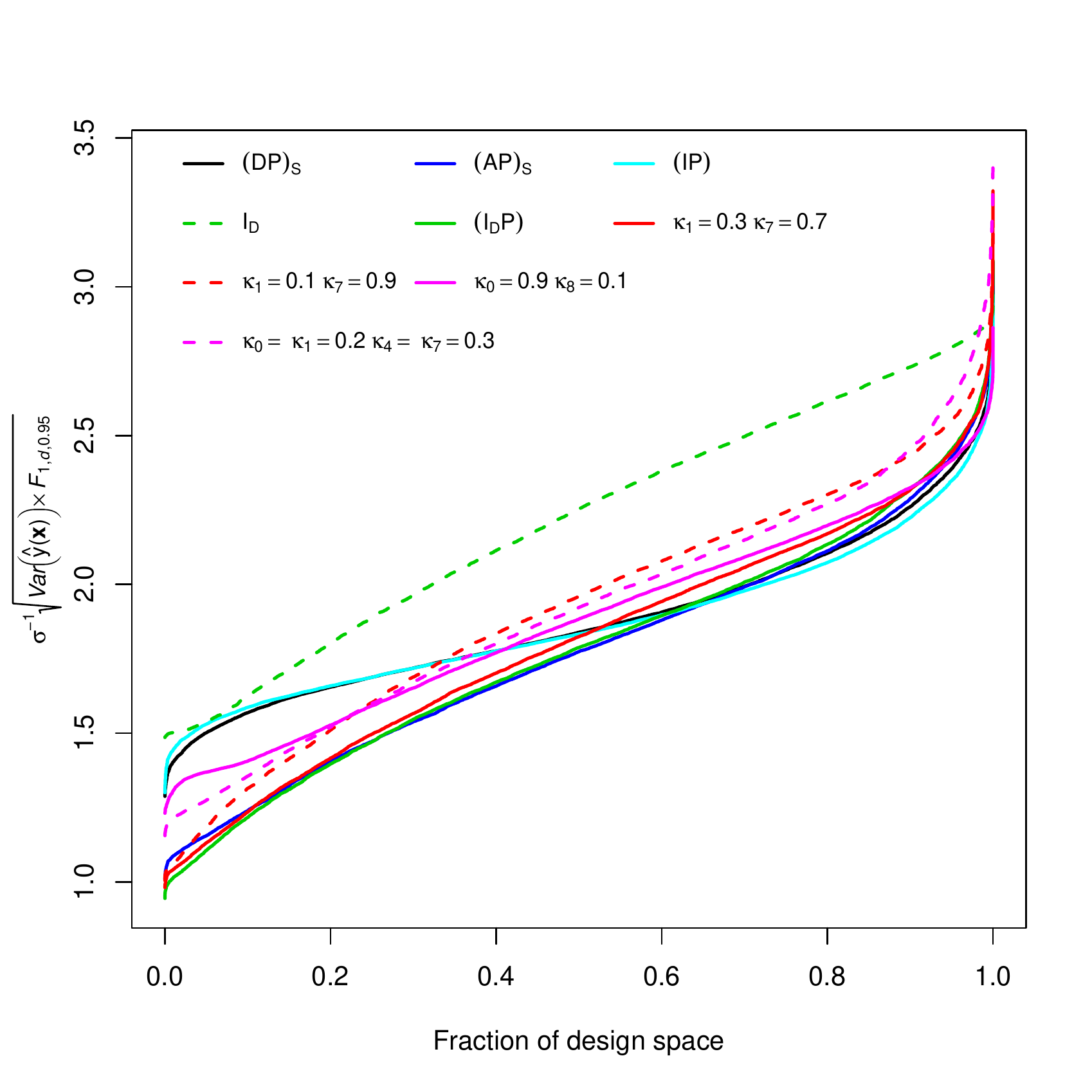}\includegraphics{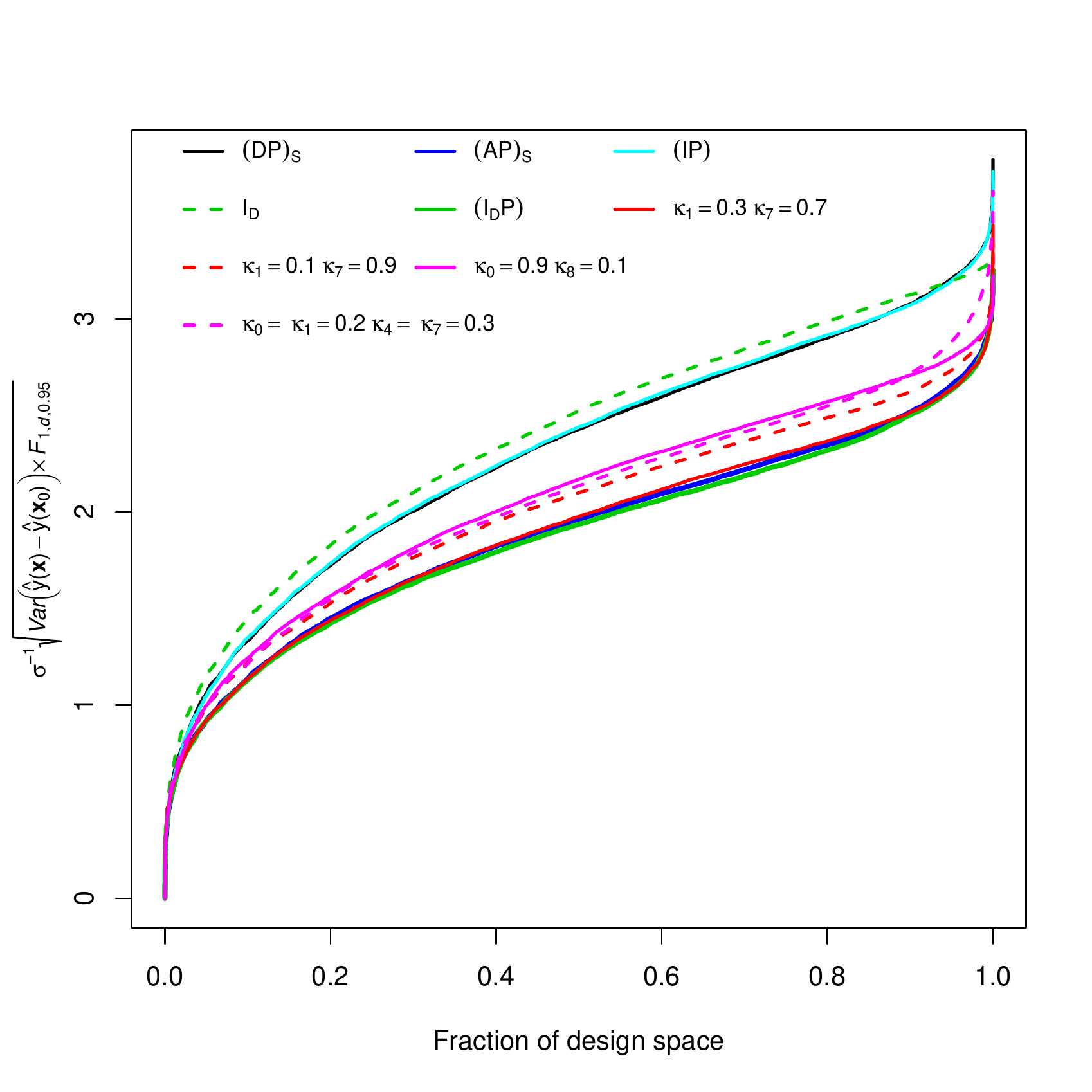}}
		\caption{FDS plots, in terms of s.e., for designs in Example 2.
			Left: response interval prediction. Right: difference interval prediction.}\label{graph:fdspeEx2}
	\end{center}
\end{figure}

In Figures \ref{graph:vdgpeEx2}-\ref{graph:fdspeEx2} (and Figures D-F in the Suppl.) we show the prediction performances of the designs over the unit hypersphere. The $D_S/I$- and $A_S$-optimum designs are not shown in the graphs referring to interval predictions because they are too poor for pure error degrees of freedom. Again we see that plotting the information against relative volume discriminates better between the designs. For response point prediction the $I_D$-, $(AP)_S$, $(I_DP)$- and compound optimum designs (8, 9, 10 and 11) have much smaller s.e.'s at the design center. However most of these designs become quite unstable away from the center.  From these, the $I_D$-optimum design is the most stable followed by design 10 (left hand-side of Figure D). Similar behavior is observed for interval response prediction (left hand-side of Figure \ref{graph:vdgpeEx2}) although $I_D$ has poorer performances than before due to few pure error degrees of freedom. The $(DP)_S$- and $(IP)$-optimum designs have very similar behavior in both graphs with poor performances at the center of the region. Perhaps fairer comparisons are obtained from Figures  \ref{graph:vdgpeEx2} and D, both right hand-side. In these graphs we can see that the advantages of designs $I_D$, 8, 9 and 11 are not so impressive since they are superior for only about $10\%$ of the region. Still, for point response predictions, their minimum values are smaller for about $30\%$ ($I_D$) and about $50\%$ (compound designs) but, because of their instability, we resort to Figure F (left hand-side) where we see lines crossing. The $D_S$/$I$-optimum design has the smallest slope but in order to achieve that, it has higher s.e.'s than other designs such as $A_S$, $I_D$ and 10 in about $50\%$ of the region. For interval response predictions (Figure \ref{graph:vdgpeEx2} the $A_S$-optimum design (not shown in the graph) and the $I_D$ optimum design are clearly no longer competitive. The $(DP)_S$- and $(IP)$-optimum designs have the smallest slopes but have higher s.e.'s than several other designs in about $40\%$ of the region. The $(AP)_S$ and $(I_DP)$-optimum design performs quite well, followed by design 8 (Figure \ref{graph:fdspeEx2}, left).

For point predictions of response differences (Figure E, left) we can identify designs $(DP)_S$, $(IP)$ and $D_S/I$ with even the minimum s.e.'s being high with the last being very stable. All other designs show smaller minimum s.e.'s.  Again the $I_D$- and $A_S$-optimum designs are quite stable but perform badly for interval predictions (Figure \ref{graph:dvdgpeEx2}, left). The compound design 10 is perhaps attractive due to its  smaller maximum s.e.'s. Once more the patterns are much clearer in Figures \ref{graph:dvdgpeEx2} and E,  both right,  which separates better the designs. The overall performances are summarized in Figures F and \ref{graph:fdspeEx2} (right). In Figure F (right) we clearly see two groups with the $(DP)_S$-, $(IP)$- and $D_S$-optimum designs having the worst performances for the whole region. The $I_D$-optimum design has the best performance throughout showing that the single criterion $I_D$ summarizes very well the point prediction capabilities in the whole region. We note, however, there are other designs with similar performances, mainly those obtained by compound criteria, although the $A_S$- and $(AP)_S$ and $(I_DP)$-optimum designs follow closely. Now, considering interval predictions of differences (Figure \ref{graph:fdspeEx2}, right), there are three designs with very close to the best performances, namely the $(AP)_S$-, $(I_DP)$- and the compound design 8 (with weights $\kappa_1=0.3$ and $\kappa_7=0.7$, compromising between $(DP)_S$ and $I_D$). The other three compound designs are also close to these.

\section{Central composite designs which are $I_D$-optimal}\label{sec:CCD}
The classical approach to designing response surface experiments, mostly commonly using CCDs, and the optimal design approach, most commonly using $D$-optimality, are often contrasted as having quite different philosophies. It is therefore intriguing that the CCD for five factors in 30 runs, based on a resolution-V half-replicate factorial portion, with four center points, in a spherical region, is optimal under the new $I_D$ criterion. It is natural to ask whether this is true for other run sizes and for other numbers of factors.

This was explored by running our exchange algorithm for various numbers of factors and run sizes in spherical regions. Subject to there being a very small chance that the algorithm has failed to find the true optimum, we found the following.
\begin{itemize}
\item For three factors, the CCD is $I_D$-optimal for $17 \leq n \leq 20$, i.e.\ 3 to 6 center points.
\item For four factors, the CCD is $I_D$-optimal for $28 \leq n \leq 32$, i.e.\ 4 to 8 center points.
\item For five factors, the CCD, with a half-replicate of the factorial points, is $I_D$-optimal for $30 \leq n \leq 33$, i.e.\ 4 to 7 center points.
\item For six factors, the CCD, with a half-replicate of the factorial points, is $I_D$-optimal for $50 \leq n \leq 55$, i.e.\ 6 to 11 center points.
\end{itemize}

We did not explore more than six factors. For other run sizes, the CCD is suboptimal. However, for run sizes just outside the range given, the optimal design is similar to a CCD, e.g.\ having one axial point replaced by a center point for small run sizes, or repeating one factorial point for larger run sizes.

Note that these CCDs are optimal only among designs chosen from the candidate set based on the full $3^q$ design, expanded to have points on the surface of the sphere. Nonetheless, we believe this is the first time CCDs have been shown to be optimal among such a large class of designs. The result nicely links the fields of classical and optimal design.

\section{Discussion}\label{sec:disc}
We have extended the compound criterion function of \cite{gilmourtrinca2012} to allow for efficient designs in terms of predictions. We focused on two properties, prediction of responses and prediction of differences in the response. Point and interval estimation were considered for both responses and differences. 

We also proposed the use of several graphs for depicting the prediction performances of the designs. We have extended the usual graphs such as VDG,  DVDG, FDS and DFDS to take into account interval estimation.  We have illustrated the methods with two examples, one for a cuboidal and the other for a spherical experimental region of interest. The illustrations showed that the graphs add relevant information mainly if one is interested in predicting the response.

Along with many other authors, we argue that a design should have several good properties and it is important to compare several designs, under a wide range of properties, in order to choose the most appropriate one for the problem at hand. This is good practice even under a single objective optimization since usually there are many designs that are almost equivalent. Evaluating them for several other properties is of great help for discriminating between them.  

The usefulness of compound criteria is that a design can be developed according to the objectives of the research. We have illustrated compound optimum designs by combining only two properties at time but of course many properties can be studied together.  
Even though this was the case for our examples, still the resulting compound designs were quite competitive overall. We have compared a compound design with the one obtained by the multiple objective algorithm of \cite{borrotti2016}. The multiple objective design did not consider inference  
and thus our compound design showed advantages. We believe that by using compound criteria we can handle many properties of interest more easily than the multiple objective approach. The graphs proposed are helpful to depict detailed pictures of prediction capabilities of the designs. We recommend the use of the proposed variations of VDG and DVDG plots that use the relative volume instead of distance for both point and interval predictions, since these graphs discriminate better between the different designs. All varieties of FDS and DFDS plots are	good summaries that are always be useful for making a final choice of design.
 
\bigskip
\begin{center}
{\large\bf SUPPLEMENTARY MATERIAL}
\end{center}

\noindent
\textbf{SuppMatPrediction.pdf}: a pdf file containing additional graphs for the examples discussed in the paper and a small simulation study to evaluate the performances of the designs in Example 1 with respect to mean and difference response bias predictions.\\
\textbf{codePrediction.rar}: a zipped folder containing R code to obtain designs by optimizing the compound criteria proposed in the article.

\bibliographystyle{Chicago}

\bibliography{Bibliography}

\begin{thebibliography}{}

\bibitem[\protect\citeauthoryear{Anderson-Cook, Borror, and
  Montgomery}{Anderson-Cook et~al.}{2009}]{anderson-cookmontg2009}
Anderson-Cook, C.~M., C.~M. Borror, and D.~C. Montgomery (2009).
\newblock Response surface design evaluation and comparison.
\newblock {\em Journal of Statistical Planning and Inference,\/}~{\em
  139\/}(2), 629--641.

\bibitem[\protect\citeauthoryear{Borrotti, Sambo, Mylona, and Gilmour}{Borrotti
  et~al.}{2017}]{borrotti2016}
Borrotti, M., F.~Sambo, K.~Mylona, and S.~Gilmour (2017).
\newblock A multi-objective coordinate-exchange two-phase local search
  algorithm for multi-stratum experiments.
\newblock {\em Statistics and Computing,\/}~{\em 27\/}(2), 469--481.

\bibitem[\protect\citeauthoryear{Box and Draper}{Box and
  Draper}{1975}]{Box-Draper:1975}
Box, G. E.~P. and N.~R. Draper (1975).
\newblock Robust designs.
\newblock {\em Biometrika,\/}~{\em 62\/}(2), 347--352.

\bibitem[\protect\citeauthoryear{Cook and Nachtsheim}{Cook and
  Nachtsheim}{1989}]{CookNachtsheim89}
Cook, R.~D. and C.~J. Nachtsheim (1989).
\newblock Computer-aided blocking of factorial and response-surface designs.
\newblock {\em Technometrics,\/}~{\em 31\/}(3), 339--346.

\bibitem[\protect\citeauthoryear{Cox and Reid}{Cox and Reid}{2000}]{coxreid}
Cox, D.~R. and N.~Reid (2000).
\newblock {\em The Theory of the Design of Experiments.}
\newblock Chapman \& Hall/CRC.

\bibitem[\protect\citeauthoryear{da~Silva, Gilmour, and Trinca}{da~Silva
  et~al.}{2017}]{daSilvaGilmourTrinca2017}
da~Silva, M.~A., S.~G. Gilmour, and L.~A. Trinca (2017).
\newblock Factorial and response surface designs robust to missing
  observations.
\newblock {\em Computational Statistics and Data Analysis,\/}~{\em 113},
  261--271.

\bibitem[\protect\citeauthoryear{Escouto}{Escouto}{2000}]{esc}
Escouto, L. F.~S. (2000, July).
\newblock {\em Desenvolvimento de produto panific\'avel a base de produtos de
  mandioca visando os hipersens\'{i}veis ao gl\'uten}.
\newblock Ph.\ D. thesis, Faculdade de Ci\^encias Agron\^omicas, Universidade
  Estadual Paulista, Botucatu.

\bibitem[\protect\citeauthoryear{Gilmour and Trinca}{Gilmour and
  Trinca}{2012}]{gilmourtrinca2012}
Gilmour, S.~G. and L.~A. Trinca (2012).
\newblock Optimum design of experiments for statistical inference (with
  discussion).
\newblock {\em Applied Statistics,\/}~{\em 61\/}(3), 345--401.

\bibitem[\protect\citeauthoryear{Giovannitti-Jensen and
  Myers}{Giovannitti-Jensen and Myers}{1989}]{g-j&m1989}
Giovannitti-Jensen, A. and R.~H. Myers (1989).
\newblock Graphical assessment of the prediction capability of response surface
  designs.
\newblock {\em Technometrics,\/}~{\em 31\/}(2), 159--171.

\bibitem[\protect\citeauthoryear{Goos and Jones}{Goos and
  Jones}{2011}]{goosjones2011book}
Goos, P. and B.~Jones (2011).
\newblock {\em Design of Experiments: a Case-Study Approach.}
\newblock Wiley.

\bibitem[\protect\citeauthoryear{Goos, Kobilinsky, O'Brien, and
  Vandebroek}{Goos et~al.}{2005}]{goosetal2005}
Goos, P., A.~Kobilinsky, T.~E. O'Brien, and M.~Vandebroek (2005).
\newblock Model-robust and model-sensitive designs.
\newblock {\em Computational Statistics and Data Analysis,\/}~{\em 49},
  201--216.

\bibitem[\protect\citeauthoryear{Hardin and Sloane}{Hardin and
  Sloane}{1991a}]{hardin1991sphere}
Hardin, R.~H. and N.~J.~A. Sloane (1991a).
\newblock Computer-generated minimal (and large) response-surface designs:
  ({I}) the sphere.
\newblock Statistics research report, AT \& T Bell Laboratories, Murray Hill,
  NJ.

\bibitem[\protect\citeauthoryear{Hardin and Sloane}{Hardin and
  Sloane}{1991b}]{hardin1991cube}
Hardin, R.~H. and N.~J.~A. Sloane (1991b).
\newblock Computer-generated minimal (and large) response-surface designs:
  ({II}) the cube.
\newblock Statistics research report, AT \& T Bell Laboratories, Murray Hill,
  NJ.

\bibitem[\protect\citeauthoryear{Hardin and Sloane}{Hardin and
  Sloane}{1993}]{hardin1993}
Hardin, R.~H. and N.~J.~A. Sloane (1993).
\newblock A new approach to the construction of optimal designs.
\newblock {\em Journal of Statistics Planning and Inference,\/}~{\em 37},
  339--369.

\bibitem[\protect\citeauthoryear{Hinkelmann and Kempthorne}{Hinkelmann and
  Kempthorne}{2008}]{hinkelmann}
Hinkelmann, K. and O.~Kempthorne (2008).
\newblock {\em Design and Analysis of Experiments.\/} (2nd ed.), Volume~1.
\newblock Wiley.

\bibitem[\protect\citeauthoryear{Jang, Anderson-Cook, and Kim}{Jang
  et~al.}{2012}]{jang2012}
Jang, D., C.~M. Anderson-Cook, and Y.~Kim (2012).
\newblock Three-dimensional quantile plots and dynamic quantile plots of
  prediction variance for response surface designs.
\newblock {\em Quality and Reliability Engineering International,\/}~{\em
  28\/}(7), 713--723.

\bibitem[\protect\citeauthoryear{Jones and Goos}{Jones and
  Goos}{2012}]{jonesgoos2012}
Jones, B. and P.~Goos (2012).
\newblock I-optimal versus {D}-optimal split-plot response surface designs.
\newblock {\em Journal of Quality Technology,\/}~{\em 44\/}(2), 85--101.

\bibitem[\protect\citeauthoryear{Jones and Nachtsheim}{Jones and
  Nachtsheim}{2011}]{jonesnash2011}
Jones, B. and C.~J. Nachtsheim (2011).
\newblock Efficient designs with minimal aliasing.
\newblock {\em Technometrics,\/}~{\em 53\/}(1), 62--71.

\bibitem[\protect\citeauthoryear{Khuri, Kim, and Um}{Khuri
  et~al.}{1996}]{khuri96}
Khuri, A.~I., H.~J. Kim, and Y.~Um (1996).
\newblock Quantile plots of the prediction variance for response surface
  designs.
\newblock {\em Computational Statistics and Data Analysis,\/}~{\em 22},
  395--407.

\bibitem[\protect\citeauthoryear{Lu, Anderson-Cook, and Robinson}{Lu
  et~al.}{2011}]{luetal11}
Lu, L., A.~C. Anderson-Cook, and T.~J. Robinson (2011).
\newblock Optimization of designed experiments based on multiple criteria
  utilizing a {P}areto frontier.
\newblock {\em Technometrics,\/}~{\em 53\/}(4), 353--365.

\bibitem[\protect\citeauthoryear{Myers, Vining, Giovannitti-Jensen, and
  Myers}{Myers et~al.}{1992}]{myers1992}
Myers, R., G.~G. Vining, A.~Giovannitti-Jensen, and S.~Myers (1992).
\newblock Variance dispersion properties of second-order response surface
  designs.
\newblock {\em Journal of Quality Technology,\/}~{\em 24\/}(1), 1--11.

\bibitem[\protect\citeauthoryear{Oliveira}{Oliveira}{2014}]{oliveira_cesar2014}
Oliveira, C. B.~A. (2014).
\newblock {\em Ferramenta computacional para avalia\c c\~ao da capacidade
  preditiva de delineamentos experimentais.}
\newblock Monografia (Gradua\c{c}\~{a}o) - Ins\-tituto de Bioci\^{e}ncias,
  Universidade Estadual Paulista, Botucatu. Available at
  \texttt{https://repositorio.unesp.br/bitstream/handle/11449/142916/000867062.pdf?\\sequence=1}.

\bibitem[\protect\citeauthoryear{Smucker and Drew}{Smucker and
  Drew}{2015}]{smucker2015}
Smucker, B. and N.~M. Drew (2015).
\newblock Approximate model spaces for model-robust experiment design.
\newblock {\em Technometrics,\/}~{\em 57\/}(1), 54--63.

\bibitem[\protect\citeauthoryear{Smucker, del Castillo, and
  Rosenberger}{Smucker et~al.}{2012}]{smucker2012}
Smucker, B.~J., E.~del Castillo, and J.~L. Rosenberger (2012).
\newblock Model-robust two-level designs using coordinate exchange algorithms
  and a maximin criterion.
\newblock {\em Technometrics,\/}~{\em 54\/}(4), 367--375.

\bibitem[\protect\citeauthoryear{Trinca and Gilmour}{Trinca and
  Gilmour}{1999}]{trincagilmour1999}
Trinca, L.~A. and S.~G. Gilmour (1999).
\newblock Difference variance dispersion graphs for comparing response surface
  designs with applications in food technology.
\newblock {\em Applied Statistics,\/}~{\em 48\/}(4), 441--455.

\bibitem[\protect\citeauthoryear{Trinca and Gilmour}{Trinca and
  Gilmour}{2017}]{trincagilmour2017}
Trinca, L.~A. and S.~G. Gilmour (2017).
\newblock Split-plot and multi-stratum designs for statistical inference.
\newblock {\em Technometrics,\/}~{\em 59\/}(4), 446--457.

\bibitem[\protect\citeauthoryear{Vining}{Vining}{1993}]{vining1993}
Vining, G.~G. (1993).
\newblock A computer program for generating variance dispersion graphs.
\newblock {\em Journal of Quality Technology,\/}~{\em 25\/}(1), 45--58.

\bibitem[\protect\citeauthoryear{Zahran, Anderson-Cook, and Myers}{Zahran
  et~al.}{2003}]{zahran2003}
Zahran, A., C.~M. Anderson-Cook, and R.~H. Myers (2003).
\newblock Fraction of design space to assess prediction capability of response
  surface designs.
\newblock {\em Journal of Quality Technology,\/}~{\em 35\/}(4), 377--386.

\end{thebibliography}

  \end{document}